\documentclass[pdflatex,sn-mathphys-num]{sn-jnl}

\usepackage{graphicx}%
\usepackage{multirow}%
\usepackage{amsmath,amssymb,amsfonts}%
\usepackage{amsthm}%
\usepackage{mathrsfs}%
\usepackage[title]{appendix}%
\usepackage{xcolor}%
\usepackage{textcomp}%
\usepackage{manyfoot}%
\usepackage{booktabs}%
\usepackage{algorithm}%
\usepackage{algorithmicx}%
\usepackage{algpseudocode}%
\usepackage{listings}%
\usepackage{dcolumn}
\usepackage{bm}
\usepackage{hyperref}
\usepackage{array}

\theoremstyle{thmstyleone}%
\theoremstyle{thmstyletwo}%

\theoremstyle{thmstylethree}%

\begin{document}

\title[Article Title]{Background-dependent and classical correspondences between 
$f(Q)$ and $f(T)$ gravity}


\author[1,2]{\fnm{Cheng} \sur{Wu}}\email{wucheng1018@mail.ustc.edu.cn}

\author[1,2,3]{\fnm{Xin} \sur{Ren}}\email{rx76@ustc.edu.cn}

\author[1,2]{\fnm{Yuhang} \sur{Yang}}\email{yyh1024@mail.ustc.edu.cn}

\author[1,2]{\fnm{Yu-Min} \sur{Hu}}\email{yumin28@ustc.edu.cn}

\author[2,4,5]{\fnm{Emmanuel N.} \sur{Saridakis}}\email{msaridak@noa.gr}

\affil[1]{Department of Astronomy, School of Physical Sciences,  
University of Science and Technology of China, 96 Jinzhai Road, Hefei, Anhui 230026, China}

\affil[2]{CAS Key Laboratory for Research in Galaxies and Cosmology, School of Astronomy and Space Science, University of Science and Technology of China, 96 Jinzhai Road, Hefei, Anhui 230026, China}

\affil[3]{Department of Physics,  Institute of Science Tokyo, Tokyo 152-8551, Japan}

\affil[4]{National Observatory of Athens, Lofos Nymfon, 11852 Athens, 
Greece}

\affil[5]{Departamento de Matem\'{a}ticas, Universidad Cat\'{o}lica del 
Norte,  Avda. Angamos 0610, Casilla 1280 Antofagasta, Chile}


\abstract{$f(Q)$ and $f(T)$ gravity are based on  fundamentally different 
geometric frameworks, yet they exhibit many similar properties. This article provides a comprehensive summary and comparative analysis of the various theoretical branches of torsional gravity and non-metric gravity, which arise from different choices of affine connection. We identify two types of background-dependent and classical correspondences 
between these two theories of gravity. The first correspondence is established through their equivalence within the Minkowski spacetime background. To achieve this, we develop the tetrad-spin formulation of $f(Q)$ gravity and derive the corresponding expression for the spin connection. The second correspondence is based on the equivalence of their equations of motion. Utilizing a metric-affine approach, we derive the general affine connection for static and spherically symmetric spacetime in $f(Q)$ gravity and compare its equations of motion with those of $f(T)$ gravity. Among others, our results reveal that,  $f(T)$ solutions are not simply a subset of $f(Q)$ solutions; rather, they encompass a complex solution beyond $f(Q)$ gravity in black hole background.}

\maketitle


\section{Introduction}
\label{sec:intro}

Modified gravity  theories offer a unique perspective on understanding the two 
phases of the Universe's accelerated expansion and provide insight into the 
physics beyond the standard cosmological model \cite{CANTATA:2021asi, 
Capozziello:2011et, Nojiri:2010wj, Carroll:2003wy, Nojiri:2003ft, 
Copeland:2006wr}. In the mathematical framework of metric-affine geometry \cite{Hehl:1994ue}, a 
prominent branch of modified gravity focuses on the geometrical trinity 
\cite{BeltranJimenez:2019esp, Heisenberg:2018vsk,Capozziello:2025hyw, Mancini:2025asp}, curvature $R$ for general 
relativity (GR), torsion $T$ for teleparallel gravity (TG), and non-metricity 
$Q$ for symmetric teleparallel gravity (STG). Since the difference between $R$ 
and $T$ (or $Q$) is merely a boundary term, the interplay of these three components 
results in two equivalent formulations of GR: the Teleparallel Equivalent of 
General Relativity (TEGR) and the Symmetric Teleparallel Equivalent of General 
Relativity (STEGR) \cite{aldrovandi2012teleparallel, Maluf:2013gaa, 
Nester:1998mp}.

While these two formulations can only yield GR-equivalent  solutions, the most 
straightforward and natural approach to obtain beyond-GR solutions is to apply a 
non-linear extension to the corresponding Lagrangian in various ways, leading to 
$f(T)$ gravity \cite{Cai:2015emx, Krssak:2018ywd}, $f(T,B)$ gravity 
\cite{Bahamonde:2015zma, Karpathopoulos:2017arc, Boehmer:2021aji, 
Bahamonde:2021gfp}, $f(Q)$ gravity \cite{ BeltranJimenez:2018vdo, 
Heisenberg:2023lru}, $f(Q,C)$ gravity \cite{De:2023xua, Capozziello:2023vne}, 
etc. These non-linear extensions have gained significant popularity in recent 
years and have been extensively explored in cosmological applications 
\cite{Clifton:2006jh, Linder:2010py, Cai:2011tc, Cai:2019bdh, Cai:2018rzd, Yan:2019gbw, 
Ren:2021tfi, Ren:2022aeo,Yang:2024kdo, Zheng:2010am, Bengochea:2010sg, 
Cardone:2012xq, Bejarano:2017akj, Golovnev:2020aon, Mavromatos:2021hai, 
Aljaf:2022fbk, Otalora:2014aoa,Chen:2010va, Escamilla-Rivera:2019ulu, 
Caruana:2020szx, Capozziello:2022dle, Kadam:2022lxt, BeltranJimenez:2019tme, 
Golovnev:2020las, Mandal:2020buf, Golovnev:2020nln, Dimakis:2021gby, 
Barros:2020bgg, Li:2021mdp, Khyllep:2021pcu, Lymperis:2022oyo, Narawade:2023tnn, 
Narawade:2023rip, Dimakis:2023uib, Lu:2019hra, De:2022wmj, Kar:2021juu, 
Mandal:2021bpd, Bahamonde:2022cmz, Capozziello:2022tvv, Hu:2023juh, 
Blixt:2023kyr, Gadbail:2023mvu, Jensko:2024bee, Jensko:2023lmn,Capozziello:2024lsz}. Furthermore, these theories have also led to interesting phenomenology in the black-hole background \cite{Hohmann:2019nat, 
Meng:2011ne, Iorio:2012cm, Xie:2013vua, Iorio:2016sqy, Farrugia:2016xcw, 
Chen:2019ftv, Ren:2021uqb, Golovnev:2021htv, DeBenedictis:2022sja, Zhao:2022gxl, 
Wang:2023qfm, Wang:2024eai, Nashed:2021pah, Ruggiero:2015oka, Nashed:2014sea,  
Bahamonde:2020bbc, Farrugia:2020fcu,  Bahamonde:2022esv, Bahamonde:2022zgj, 
Lin:2021uqa, Hohmann:2019fvf, DeFalco:2023twb, Banerjee:2021mqk, Awad:2022fhx,   
         Junior:2023qaq, Javed:2023vmb, Gogoi:2023kjt, Wang:2021zaz, 
Junior:2024xmm, Calza:2022mwt, Calza:2023hhi}. 

In addition to their  cosmological and black hole applications, the connection 
branches of $f(T)$ and $f(Q)$ gravity in different backgrounds, derived through 
symmetry analysis, have become an increasingly popular topic in recent studies 
\cite{Bahamonde:2021srr, Hohmann:2021ast, DAmbrosio:2021pnd, DAmbrosio:2021zpm,  
Dimakis:2022rkd, Shi:2023kvu, Gomes:2023hyk, Subramaniam:2023okn, Yang:2024tkw, 
Guzman:2024cwa}. In the case of the static and spherically symmetric spacetime 
within $f(T)$ gravity, three tetrads in the Weitzenböck gauge correspond to 
three distinct branches of solutions \cite{Bahamonde:2021srr}. Meanwhile, the 
static and spherically symmetric spacetime of $f(Q)$ gravity was discussed in 
\cite{DAmbrosio:2021zpm}, where the authors summarized different sets of 
constraint equations of the affine connection and highlighted that black hole 
solutions in $f(T)$ gravity are merely a subset of those in $f(Q)$ gravity. In 
the cosmological spacetime with zero spatial curvature, $f(T)$ 
gravity has only one branch \cite{Hohmann:2019nat, DAmbrosio:2021pnd}, whereas 
$f(Q)$ gravity has three branches \cite{Hohmann:2021ast, DAmbrosio:2021pnd}. 

To understand why $f(Q)$ and $f(T)$ gravity  have different branches  in the 
same background, it is important to note that the usual formulations of TG and 
STG are different; TG is based on the tetrad-spin formulation, while STG relies 
on the metric-affine formulation \cite{Krssak:2018ywd, Xu:2019sbp}.
Although those two formulations are equivalent,  the distinct geometric 
backgrounds affect which formulation is more convenient for different gravity 
theories. Furthermore, variations in parameterizations between the two 
formulations can yield different solutions based on their respective parameter 
spaces. To understand these multiple branches and their correspondences, we 
argue that it is essential to use the tetrad-spin formulation to describe STG. 
This approach is primarily used in TG, through which a complex solution  has 
been discovered \cite{Bahamonde:2021srr}. Additionally, in both TG and STG, 
there is a method to derive an appropriate form of the spin connection or affine 
connection by switching off gravity, providing a unique perspective to 
understand the correspondence between these two theories \cite{Krssak:2015oua, 
Zhao:2021zab}.

The aim of this article is to establish correspondences  between different 
connection branches in $f(Q)$ and $f(T)$ gravity. Typically, there are two 
approaches to derive the form of the connection in these theories: one is by 
switching off gravity, while the other relies on symmetry analysis. Accordingly, 
it is natural to propose two distinct correspondences based on these approaches: Minkowski-equivalence (ME) correspondence and equations-of-motion (EoMs) correspondence. 
However, both correspondences are background-dependent, as the connection 
branches are determined only within specific backgrounds.

The outline of this article is as follows.  In Section~\ref{sec:definitions}, 
we provide a brief review of geometrical trinity and flat gravity theories in 
their preferred formulations. In Section~\ref{sec:connection_branches}, we summarize different branches of $f(Q)$ and $f(T)$ gravity in different backgrounds. In Section~\ref{sec:Minkowski_eq_correspondence}, we develop the 
tetrad-spin formulation of $f(Q)$ gravity, calculate the field equations 
within this framework, and then establish the Minkowski-equivalence
correspondence between $f(Q)$ and $f(T)$ gravity. In Section~\ref{sec:EoMs_Correspondence}, we establish the equations-of-motion correspondence between $f(Q)$ and $f(T)$ gravity. Finally, we end in Section~\ref{sec:conclusion} with the conclusions.

\section{Covariant $f(Q)$ gravity and $f(T)$ gravity }
\label{sec:definitions}

\subsection{Geometrical trinity in metric-affine and tetrad-spin formulation  }

We begin with a brief review of  the general metric-affine geometry, general 
tetrad-spin geometry, and the definition of geometrical trinity in those two 
formulations. In metric-affine theory, the metric $g_{\mu\nu}$ and affine 
connection $\Gamma^\nu{}_{\rho \mu}$ of spacetime are employed to describe 
gravity. While in the tetrad-spin framework, the tetrad $h^a{}_{\mu}$ and spin 
connection ${A^a}_{b\mu}$ are utilized. Note that these two approaches are 
merely different depictions of gravity, and the ultimate physics remains the 
same.

We adopt the convention in which the last index of the connection serves as  
the ``derivative index'', namely $\nabla_\mu V^\nu = \partial_\mu 
V^\nu+\Gamma^\nu{}_{\rho \mu}V^\rho$. We use Greek letters ($\mu, \nu, ...$) to 
denote coordinate indices and Latin letters ($a, b, ...$) for tangent space 
indices. 

We begin with the metric-affine formulation, the metric tensor is  denoted by 
$g_{\mu \nu}$ and the covariant derivative associated with the affine connection 
$\Gamma^{\lambda}{}_{\mu\nu}$ is given by:
\begin{align}
    \nabla_\mu\phi^\nu=\partial_\mu \phi^\nu+{\Gamma^\nu}_{\rho\mu}\phi^\rho,\\
    \nabla_\mu\phi_\nu=\partial_\mu \phi_\nu-{\Gamma^\rho}_{\nu\mu}\phi_\rho.
\end{align}
Under a coordinate transformation $\{x^\mu\}\rightarrow \{x'^{\mu}\}$, in order 
to maintain the  covariance of the covariant derivative, the affine connection 
transforms as:
\begin{equation}
    {\Gamma'^\rho}_{\mu \nu}=\frac{\partial x'^\rho}{\partial x^\tau} \frac{\partial x^\omega}{\partial x'^\mu} \frac{\partial x^\sigma}{\partial x'^\nu}{\Gamma^\tau}_{\omega \sigma}+\frac{\partial x'^\rho}{\partial x^\sigma}\frac{\partial^2 x^\sigma}{\partial x'^\nu \partial x'^\mu}.
\end{equation}
The geometrical trinity, namely the curvature tensor, the torsion tensor and the
non-metricity tensor, in the metric-affine formulation are defined as
\begin{align}
    {R^\rho}_{\lambda \nu \mu} &\equiv \partial_\nu {\Gamma^\rho}_{\lambda \mu}-\partial_\mu {\Gamma^\rho}_{\lambda \nu}+{\Gamma^\rho}_{\eta \nu}{\Gamma^\eta}_{\lambda \mu}-{\Gamma^\rho}_{\eta \mu}{\Gamma^\eta}_{\lambda \nu},\\
    {T^\rho}_{\nu \mu} &\equiv {\Gamma^\rho}_{\mu \nu}-{\Gamma^\rho}_{\nu \mu},\\
    Q_{\alpha\mu\nu} &\equiv \nabla_\alpha g_{\mu \nu} = \partial_\alpha g_{\mu \nu}-{\Gamma^\lambda}_{\mu \alpha}g_{\lambda \nu}-{\Gamma^\lambda}_{\nu \alpha}g_{\mu\lambda}. \label{eq: non-metricity tensor definition}
\end{align}
Applying Eq.~\eqref{eq: non-metricity tensor definition} and permutating the 
indices, we obtain the decomposition of the affine connection as
\begin{align}
    {\Gamma^\rho}_{\mu \nu} &= \{ \begin{array}{c} \rho\\ \mu \,\nu \end{array} \}+{K^\rho}_{\mu \nu}+{L^\rho}_{\mu \nu}, \label{decomposition of affine connection}
\end{align}
where $\{ \begin{array}{c} \rho\\ \mu \,\nu \end{array} \}$ is the Christoffel 
symbol, $K^\rho{}_{\mu\nu}$ is the contortion tensor and $L^\rho{}_{\mu\nu}$ is 
the disformation tensor:
\begin{align}
    \{ \begin{array}{c} \rho\\ \mu \,\nu \end{array} \} &\equiv 
\frac{1}{2}g^{\rho \sigma}(\partial_\nu g_{\mu \sigma}+\partial_\mu g_{\nu 
\sigma}-\partial_\sigma g_{\mu \nu}),\\ 
    {K^\rho}_{\mu \nu} &\equiv 
\frac{1}{2}(T_\mu{}^\rho{}_\nu+T_\nu{}^\rho{}_\mu-T^\rho{}_\mu{}_\nu),\\
    {L^\rho}_{\mu \nu} &\equiv \frac{1}{2}({Q^\rho}_{\mu 
\nu}-Q_\mu{}^\rho{}_\nu-Q_\nu{}^\rho{}_\mu).
\end{align}

We proceed to the tetrad-spin formulation. The metric tensor 
$g_{\mu\nu}$ and the tetrad field $h^a{}_{\mu}$ are related by 
\begin{equation}
    g_{\mu\nu} = h^a{}_\mu h^b{}_{\nu} \eta_{ab}, \label{eq: tetrad definition}
\end{equation}
where  $\eta_{ab}$ is the Minkowski metric. 

The covariant derivative associated with the spin connection ${A^a}_{b\mu}$ is 
given by: 
\begin{align}
    \mathcal{D}_\mu\phi^c=\partial_\mu \phi^c+{A^c}_{d\mu}\phi^d,\\
    \mathcal{D}_\mu\phi_c=\partial_\mu \phi_c-{A^d}_{c\mu}\phi_d. \label{2.5}
\end{align}
 Additionally, we assume that the tetrad satisfies the following identity, known 
as the "tetrad postulate"  \cite{aldrovandi2012teleparallel}:
\begin{equation}
     \partial_\mu h^a{}_{\nu}+A^a{}_{b \mu} h^b{}_\nu-\Gamma^\rho{}_{\nu 
\mu}h^a{}_\rho \equiv 0.
    \label{eq: tetrad postulate}
\end{equation}
From the tetrad postulate, we can establish the relationship between the spin 
connection and the affine connection as
\begin{align}   
{\Gamma^\rho}_{\nu\mu}={h_a}^{\rho}\partial_\mu{h^a}_\nu+{h_a}^\rho{A^a}_{b\mu}{
h^b}_\nu={h_a}^\rho\mathcal{D}_\mu{h^a}_\nu, \label{eq: affine connection 
definition}\\
 {A^a}_{b\mu}={h^a}_{\nu}\partial_\mu{h_b}^\nu+{h^a}_\nu{\Gamma^\nu}_{\rho\mu}{ 
h_b}^\rho={h^a}_\nu\nabla_\mu{h_b}^\nu. \label{eq: spin connection definition}
\end{align}

Under a tetrad transformation $h^a{}_\mu \rightarrow h'^a{}_\mu=\Lambda^a{}_b 
h^b{}_\mu$ (where ${\Lambda^a}_b$ are components belonging to a Lorentz group), 
the spin connection transforms as
\begin{equation}    
{A'^a}_{b\mu}={\Lambda^a}_c{\Lambda_b}^d{A^c}_{d\mu}+{\Lambda^a}_c\partial_\mu{
\Lambda_b}^c.
\end{equation}
Combining with Eq.~\eqref{eq: tetrad definition}, Eq.~\eqref{eq: affine 
connection definition} and Eq.~\eqref{eq: spin connection definition}, we can 
derive the definition of geometrical trinity in the tetrad-spin formulation, 
namely
\begin{align}   
    {R^a}_{b \mu \nu} &=\partial_\nu {A^a}_{b \mu}-\partial_\mu {A^a}_{b 
\nu}+{A^a}_{e \nu}{A^e}_{b \mu}-{A^a}_{e \mu}{A^e}_{b \nu}, \label{eq: curvature 
tetrad-spin}\\
    {T^a}_{\nu \mu}&=\partial_\nu{h^a}_\mu-\partial_\mu 
{h^a}_\nu+{A^a}_{e\nu}{h^e}_\mu-{A^a}_{e \mu}{h^e}_\nu,\\
   Q_{\lambda a b} &= -\eta_{ac}A^c{}_{b\lambda}-\eta_{bc}A^c{}_{a\lambda}. 
\label{eq: non-metricity tetrad-spin}
\end{align}
The coefficient of anholonomy is defined by:
\begin{equation}
    {f^c}_{ab}={h_a}^\mu {h_b}^\nu(\partial_\nu{h^c}_\mu-\partial_\mu{h^c}_\nu), 
\label{eq: anholonomy coefficient}
\end{equation}
which represents the non-commutativity of tetrad.  If $f^c{}_{ab}=0$ then we 
state that the tetrad is holonomic. Using Eq.~\eqref{eq: anholonomy 
coefficient}, we can find the relationship between the torsion tensor and the 
spin connection as
\begin{align}
    A^a{}_{cb}-A^a{}_{bc}=T^a{}_{bc}+f^a{}_{bc}.
\end{align}
By permutation of indices, we derive the decomposition of the spin connection:
\begin{align}
    A_{abc}&=A_{[ab]c}+A_{(ab)c}\notag\\
&=\mathring{\omega}_{abc}+K_{abc}+L_{abc},
\end{align}
where $\mathring{\omega}^a{}_{bc}$ is the spin connection in general 
relativity, $K^a{}_{bc}$ is the contortion tensor and $L^a{}_{bc}$ is the 
disformation tensor:
\begin{align}
    \mathring{\omega}^a{}_{bc} &\equiv \frac{1}{2}(f_b{}^a{}_c+f_c{}^a{}_b-f^a{}_{bc}),\\
    {K^a}_{bc} &\equiv \frac{1}{2}(T_b{}^a{}_c+T_c{}^a{}_b-T^a{}_b{}_c),\\
    {L^a}_{b c} &\equiv \frac{1}{2}({Q^a}_{b c}-Q_b{}^a{}_c-Q_c{}^a{}_b).
\end{align}

\subsection{$f(Q)$ gravity and $f(T)$ gravity in their preferred formulations}

In this section, we compare $f(Q)$ gravity and $f(T)$ gravity in their preferred 
 formulations. Despite being rooted in different geometric frameworks, these 
theories exhibit numerous similarities.

\subsubsection{Metric-affine formulation of $f(Q)$ gravity}

In teleparallel geometry, the flat condition requires vanishing curvature, thus the resulting affine connection can be given by
\begin{equation}
    {\Gamma^\alpha}_{\mu \nu} = {(M^{-1})^{\alpha}}_{\lambda} \partial_\nu {M^{\lambda}}_{\mu},
\end{equation}
where $M^\mu{}_\nu$ are components of a matrix belonging to the general linear group $GL(4, \mathbb{R})$ \cite{Heisenberg:2023lru}. For symmetric teleparallel gravity, the torsionless condition further restricts the affine connection to the form:
\begin{equation}
    {\Gamma^\alpha}_{\mu \nu}=\frac{\partial x^\alpha}{\partial  
\xi^\lambda}\partial_\nu \partial_\mu \xi^\lambda,
    \label{eq: coincident affine connection}
\end{equation}
where $\xi^\mu$ is an arbitrary function and is used to parametrize the affine 
connection. Under a special gauge fixing on coordinates by $ 
\{x^\mu\}\rightarrow \{\xi^\mu\}$, which is referred as the coincident gauge and 
is always available, the affine connection at all points vanishes 
automatically. In other words, for an arbitrary coordinate system, 
the coincident gauge can be achieved through an appropriate coordinate 
transformation. Additionally, $\{\xi^\mu\}$ can also be referred to as 
St\"uckelberg fields since the definition of non-metricity tensor can be 
reobtained through the St\"uckelberg formulation,  which restores diffeomorphisms by 
promoting $\partial_\alpha g_{\mu\nu}$ to a covariant object 
\cite{BeltranJimenez:2022azb}.

Furthermore, the parametrization form of the affine connection~\eqref{eq: 
coincident affine connection} indicates that the affine connection is solely 
related with the coordinate transformation, independently of   gravity. 
Therefore, in order to determine the affine connection in $f(Q)$ gravity, a 
practical way is to find the corresponding metric in Minkowski spacetime, 
namely  to remove parameters containing gravitational information in the metric 
when gravity still exists. 
By calculating the connection in Minkowski spacetime,  we obtain the affine 
connection in the case where gravity does not vanish. 
If we assume that   non-metricity is zero in Minkowski spacetime, then according 
to Eq.~\eqref{decomposition of affine connection}  the affine connection 
simplifies to the Levi-Civita connection \cite{Zhao:2021zab}.

The action of $f(Q)$ gravity is defined as
\begin{align}
    \mathcal{S}=-\frac{1}{2 \kappa}\int d^4x \, \sqrt{-g}f(Q)+\mathcal{S}_{matter},
\end{align}
where  $\kappa=8\pi G$, $g=\det(g_{\mu \nu})$ and $\mathcal{S}_{matter}=\int 
d^4x \, \mathcal{L}_{matter}$ represents the action of matter fields. In the 
above expression, we have defined the non-metricity scalar as:
\begin{align}
  Q \equiv \frac{1}{4}Q_{\alpha \mu \nu}Q^{\alpha \mu \nu}-\frac{1}{2}Q_{\alpha \mu \nu}Q^{\mu \alpha \nu}-\frac{1}{4}Q_\alpha Q^\alpha+\frac{1}{2}Q_\alpha \bar{Q}^\alpha,
\end{align}
where $Q_\alpha \equiv g^{\mu \nu} Q_{\alpha \mu \nu}$ and $\bar{Q}\equiv g^{\mu \nu} Q_{\mu \alpha \nu}$.
Performing variation of the action with respect to the metric tensor and 
the affine connection, we obtain the field equations of $f(Q)$ gravity, namely
\begin{align}
    E^{\mu\nu} \equiv \frac{1}{\sqrt{-g}}\nabla_\alpha(\sqrt{-g}f_QP^{\alpha 
\mu \nu})+f_Q({P^{\alpha \beta(\mu}}{Q_{\alpha 
\beta}}^{\nu)}+\frac{1}{2}{P^{(\nu}}_{\alpha \beta}Q^{\mu) \alpha 
\beta})+\frac{1}{2}f\,g^{\mu\nu} = \kappa \mathcal{T}^{\mu \nu}, \label{eq: f(Q) 
metric EoM}\\
   2\nabla_\nu \nabla_\mu(\sqrt{-g}f_Q{P^{\mu 
\nu}}_\lambda)=\nabla_\nu\nabla_\mu \mathcal {H_\lambda}^{\nu\mu}, \label{eq: 
f(Q) affine EoM}
\end{align}
where $f_Q=\frac{df(Q)}{dQ}$. Finally, we define the non-metricity conjugate 
as 
\begin{align}
  {P^\alpha}_{\mu \nu}&=-\frac{\partial Q}{\partial {Q_\alpha}^{\mu \nu}} = 
-\frac{1}{2}{Q^\alpha}_{\mu \nu}+Q_{(\mu}{}^\alpha{}_{\nu)}+\frac{1}{2}g_{\mu 
\nu}(Q^\alpha-\bar{Q}^\alpha)-\frac{1}{2}\delta^\alpha{}_{(\mu}Q_{\nu)},
\end{align}
the energy-momentum tensor as
\begin{align}
    \mathcal{T}^{\mu \nu} \equiv \frac{2}{\sqrt{-g}}\frac{\delta \mathcal{L}_{matter}}{\delta g_{\mu \nu}},
\end{align}
and the hypermomentum tensor as
\begin{align}
   \mathcal{H}_\alpha{}^{\mu \nu} \equiv 2\kappa \frac{\delta \mathcal{L}_{matter}}{\delta {\Gamma^\alpha}_{\mu \nu}}.
\end{align}

\subsubsection{Tetrad-spin formulation of $f(T)$ gravity} 

For teleparallel gravity, the flat and metric-compatible condition constrains 
the spin connection to the form  
\begin{equation}
    A^a{}_{b\mu}={\Lambda^a}_e \partial_\mu {\Lambda_b}^e, \label{eq: Weitzenbock}
\end{equation}
where $\Lambda^a{}_b$ are components of a matrix belonging to the Lorentz group \cite{Krssak:2018ywd}.
Analogously to $f(Q)$ gravity, the spin connection in $f(T)$ gravity is solely 
related to the Lorentz transformation, independently gravity. We refer to the 
affine connection associated with Eq.~\eqref{eq: Weitzenbock} as Weitzenb\"ock 
connection.

When gravity is switched off, the spin connection retains its value and the 
tetrad can be expressed as 
\begin{align}
    {h^a}_\mu=\partial_\mu v^a+{\omega^a}_{b\mu}v^b, \label{eq:tetrad_Lorentz_vector}
\end{align}
where $\omega^a{}_{b\mu}$ is the Lorentz connection (defined as the spin connection with vanishing symmetric components) and $v^a$ is the Lorentz vector. If Lorentz connection is zero, then the tetrad in Minkowski spacetime is holonomic. 

The action of $f(T)$ gravity is defined as 
\begin{align}
    \mathcal{S}=-\frac{1}{2 \kappa}\int d^4x \, h f(T)+\mathcal{S}_{matter},
\end{align}
where $h=\det{(h^a{}_\mu)}$, and the torsion scalar is 
\begin{align}
    T \equiv \frac{1}{4}{T^\rho}_{\mu \nu}{T_\rho}^{\mu \nu}+\frac{1}{2}{T^\rho}_{\mu \nu}{T^{\nu\mu}}_\rho-T_\mu T^\mu.
\end{align}
Performing variation of the action with respect to the tetrad and the spin 
connection, we   derive the field equations of $f(T)$ gravity as 
\begin{align}
   E_a{}^\mu \equiv \frac{1}{h}f_T \partial_\nu(h{S_a}^{\mu 
\nu})+f_{TT}{S_a}^{\mu \nu} \partial_\nu T-f_T {T^b}_{\nu a}{S_b}^{\nu \mu}+f_T 
{A^b}_{a \nu}{S_b}^{\nu \mu} +\frac{1}{2}f{h_a}^\mu =\kappa{\mathcal{T}_a}^\mu, 
\label{eq: f(T) tetrad EoM}\\
    f_{TT}\,\partial_\mu T\,h{{S_{[ab]}}^{\mu}} =0,
\end{align}
where we have defined the superpotential as
\begin{align}
    {S_a}^{\rho \sigma}=\frac{1}{2}({T^{\sigma \rho}}_a+{T_a}^{\rho \sigma}-{T^{\rho \sigma}}_a)-{h_a}^\sigma T^{\rho}+{h_a}^\rho T^\sigma
\end{align}
and the energy-momentum tensor as 
\begin{align}
    {\mathcal{T}_a}^\mu \equiv \frac{1}{h}\frac{\delta \mathcal{L}_{matter}}{\delta {h^a}_\mu}.
\end{align}

\section{Connection branches in teleparallel gravity theories}
\label{sec:connection_branches}
In this section, we summarize the connection branches of $f(Q)$ and $f(T)$ in cosmological and black hole spacetime.

\subsection{Cosmological background}
The metric and tetrad in cosmological spacetime are chosen as
\begin{align}
    g_{\mu\nu} &= diag\{-1, a(t)^2, a(t)^2r^2, a(t)^2r^2\sin^2\theta\},\\
    {h^a}_\mu &= diag\{1, a(t), a(t)\,r, a(t)\, r\sin\theta\}. \label{eq:tetrad_cosmological}
\end{align}

For $f(Q)$ gravity, there are three branches, which are expressed as \cite{Hohmann:2019fvf}
\begin{equation}
    \begin{aligned}
        &\Gamma^{t}_{\ tt}=C_1,\quad \Gamma^{t}_{\ rr}=C_2,\quad \Gamma^{t}_{\ 
    \theta\theta}=C_2r^2\quad \Gamma^{t}_{\ \phi\phi}=C_2r^2\sin^2{\theta}, \\
        &\Gamma^{r}_{\ tr}=C_3,\quad \Gamma^{r}_{\ rr}=0,\quad \Gamma^{r}_{\ 
    \theta\theta}=-r,\quad \Gamma^{r}_{\ \phi\phi}=-r\sin^2{\theta}, \\
        &\Gamma^{\theta}_{\ t\theta}=C_3,\quad \Gamma^{\theta}_{\ 
    r\theta}=\frac{1}{r},\quad \Gamma^{\theta}_{\ 
    \phi\phi}=-\cos{\theta}\sin{\theta},  \\
        &\Gamma^{\phi}_{\ t\phi}=C_3,\quad \Gamma^{\phi}_{\ 
    r\phi}=\frac{1}{r},\quad 
    \Gamma^{\phi}_{\ \theta\phi}=\cot{\theta},
    \end{aligned}
    \end{equation}
where $C_1$, $C_2$, $C_3$ and non-metricity scalar have three sets of choices in Table \ref{tab:connection_branches_in_cosmological_spacetime}.

Their St\"uckelberg fields are
\begin{align}
    \xi_{ \uppercase\expandafter{\romannumeral1}}&=\{\zeta(t),\zeta(t)r \sin \theta \cos \phi,\zeta(t)r \sin \theta \sin \phi,\zeta(t)r \cos \theta\},\\
    \xi_{ \uppercase\expandafter{\romannumeral2}}&=\{\zeta(t)+\frac{1}{2}r^2,r \sin \theta \cos \phi,r \sin \theta \sin \phi,r \cos \theta\},\\
    \xi_{ \uppercase\expandafter{\romannumeral3}}&=\{\zeta(t),r \sin \theta \cos \phi,r \sin \theta \sin \phi,r \cos \theta\},
\end{align}
where $\frac{\ddot\zeta}{\dot\zeta}=C_1$.

For Branch \uppercase\expandafter{\romannumeral1}, the field equations are
\begin{align}
    \frac{6 \dot a^2}{a^2} f_Q - \frac{1}{2} f = \kappa \rho, \notag\\
    -4 \dot a^2 f_Q - 2 a \left( \ddot a f_Q + \dot a \dot Q f_{QQ} \right) + \frac{1}{2} a^2 f = \kappa p.
\end{align}

For Branch \uppercase\expandafter{\romannumeral2}, the field equations are
\begin{align}
    \frac{6 \dot a^2}{a^2} f_Q-\frac{1}{2}f-\frac{1}{2} \left(-3 \gamma \dot Q f_{QQ}+3 \dot\gamma  f_Q\right) 
    -\frac{9\dot a}{2a} \gamma f_Q = \kappa \rho,  \notag \\
    -4 \dot a^2 f_Q-2 a \left(\ddot a f_Q+\dot a \dot Q f_{QQ}\right)+\frac{1}{2} a^2 f
    +  \frac{1}{2} a\left(9 \gamma \dot a f_Q+a \left(3 \gamma \dot Q f_{QQ}+3 \dot \gamma f_Q\right)\right) = \kappa p.
\end{align}

For Branch \uppercase\expandafter{\romannumeral3}, the field equations are
\begin{align}
    \frac{6 \dot a^2}{a^2}f_Q-\frac{1}{2} f -\frac{3 \gamma \dot a f_Q}{2a^3}-\frac{3 \left(\gamma  \dot Q f_{QQ}+\dot \gamma f_Q\right)}{2 a^2}= \kappa \rho, \notag\\
    -4 \dot a^2 f_Q-2 a \left(\ddot a f_Q+\dot a \dot Q f_{QQ}\right)+\frac{1}{2} a^2 f+\frac{3 \gamma \dot a f_Q}{2a}+\frac{1}{2} \left(\gamma \dot Q f_{QQ}+3 \dot \gamma f_Q\right) = \kappa p.
\end{align}
\begin{table*}
    \centering
    \caption{Different branches of f(Q) and f(T) theory in cosmological background. $G\rightarrow0$ denotes the case when gravity vanishes, namely in Minkowski spacetime with $a(t)=1$. Since the properties of $\gamma$ are unknown, the cell $-3\dot\gamma$ may not be accurate if $\gamma$ changes its value when gravity is switched off.}
    \vspace{10pt}
    \renewcommand{\arraystretch}{2}
    \begin{tabular}{ccccccc}
    \hline
    \multirow{4}{*}{$f(Q)$}& Branch &$C_1$ &$C_2$ &$C_3$ & $Q$ & $Q_{G\rightarrow 0}$\\
    \cmidrule(lr){2-7}
    &\uppercase\expandafter{\romannumeral1} &$\gamma$ &$0$ &$0$ &$\frac{6 \dot a^2}{a^2}$ &  0    \\
    \cmidrule(lr){2-7}
        &\uppercase\expandafter{\romannumeral2}
    &$\gamma+\frac{\dot{\gamma}}{\gamma}$ &$0$ &$\gamma$ &$-\frac{9 \gamma \dot a}{a}+\frac{6 \dot a^2}{a^2}-3 \dot \gamma$ & $-3\dot \gamma $\\
    \cmidrule(lr){2-7}
        & \uppercase\expandafter{\romannumeral3}
    &$-\frac{\dot{\gamma}}{\gamma}$ &$\gamma$ & $0$ & $-\frac{3 \left(a \left(\dot\gamma -2 \dot a^2\right)+\gamma \dot a\right)}{a^3}$ & $-3\dot \gamma  $\\
    \hline
    \multirow{2}{*}{$f(T)$}& \multicolumn{4}{c}{Branch} & $T$ & $T_{G\rightarrow 0}$\\
    \cmidrule(lr){2-7}
    &\multicolumn{4}{c}{ME,T}&$\frac{6{\dot a}^2}{a^2}$&0\\
    \hline
    \end{tabular}
    \label{tab:connection_branches_in_cosmological_spacetime}
\end{table*}

For $f(T)$ gravity, there is only one branch, which we refer as Minkowski-equivalence correspondence branch, "ME Branch" for short, with the torsion scalar $T=\frac{6{\dot a}^2}{a^2}$:
\begin{align}
    {\omega^r}_{\theta\theta} &= -1,\quad &{\omega^r}_{\phi\phi} &= -\sin\theta, \notag\\
    {\omega^\theta}_{r\theta} &= 1, \quad &{\omega^\theta}_{\phi\phi}&=-\cos\theta, \notag\\
    {\omega^\phi}_{r\phi} &= \sin\theta, \quad &{\omega^\phi}_{\theta\phi}&= \cos\theta. \label{eq:spin_ME}
\end{align}

The Lorentz vector in Weitzenb\"ock gauge is
\begin{equation}
    v^a = \{t, r\sin\theta\cos\phi, r\sin\theta\sin\phi, r\cos\theta\}.\label{eq:Lorentz_vector_ME}
\end{equation}

The field equations are
\begin{align}
    \frac{6 \dot a^2}{a^2}f_T-\frac{1}{2} f =\kappa \rho ,\\
    -4 \dot a^2 f_T-2 a\left(\ddot a f_T+\dot a \dot T f_{TT}\right)+\frac{1}{2}{\dot a^2} f=\kappa p.
\end{align}

\subsection{Black hole background}
The metric and tetrad in cosmological spacetime are chosen as
\begin{align}
    g_{\mu\nu} &= diag\{-A(r)^2, B(r)^2, r^2, r^2\sin^2\theta\},\\
    {h^a}_\mu &= diag\{A(r), B(r), r, r\sin\theta\}. \label{eq:tetrad_blackhole}
\end{align}

For $f(Q)$ gravity, we present three special branches while the general one is discussed in Section~\ref{sec:EoMs_Correspondence}.

The first branch $\Gamma_{ME,Q}$ is
\begin{align}
    {\Gamma^r}_{\theta\theta} &= -r, \quad &{\Gamma^r}_{\phi\phi} &= -r\sin^2\theta, \notag\\
    {\Gamma^\theta}_{r\theta} &= {\Gamma^\theta}_{\theta r} =\frac{1}{r}, \quad 
&{\Gamma^\theta}_{\phi\phi}&=-\cos\theta \sin\theta,\notag\\ 
    {\Gamma^\phi}_{r\phi} &= {\Gamma^\phi}_{\phi r}=\frac{1}{r}, \quad 
&{\Gamma^\phi}_{\theta\phi} &= {\Gamma^\phi}_{\phi \theta}=\cot\theta,\label{eq:min_affine}
\end{align}
with
\begin{align}
Q_{ME} & =\frac{2 \left(B^2-1\right) \left(B A'+A B'\right)}{r A B^3},\\
Q_{ME,G\rightarrow 0} & =0, \\
\xi_{ME}^a &= \{t, r\sin\theta\cos\phi, r\sin\theta\sin\phi, r\cos\theta\}.
\end{align}
Its EoMs are
\begin{align}
    E_{ME,00}=&\frac{A}{2 r^2 B^3}(\left(2 r A B^3 Q'-2 r A B Q'\right)f_{QQ}\notag\\
    &+(2 r \left(B^2-1\right) B A'+2 r A B^2 B'+2 r A B'+2 A B^3-2 A B) f_Q-r^2 A B^3 f),\\
    E_{ME,11}=&-\frac{1}{2r^2AB}(\left(2 r A B^3 Q'-2 r A B Q'\right)f_{QQ}\notag\\
    &+(2 r \left(B^2-3\right) B A'+2 r A B^2 B'-2 r A B'+2 A B^3-2 A B) f_Q-r^2 A B^3 f),\\
    E_{ME,22}=&\frac{E_{ME,33}}{\sin^2\theta}=\frac{r}{2 A B^3}(2 r B A' Q' f_{QQ}\notag\\
    &+\left(2 r B A''-2 r A' B'-2 B^3 A'+4 B A'-2 A B^2 B'\right)f_Q+r A B^3 f).
\end{align}
We find $E_{ME,00}-E_{ME,11}\left(-\frac{A^2}{B^2}\right)=\frac{2A}{rB^3}(BA'+AB')f_Q$. For the vacuum case, $BA'+AB'=0$ so $Q_{ME}=0$, which leading to the Schwarzschild solution \cite{Zhao:2021zab}.

The other two branches (we call them $\xi$ branch) are
\begin{align}
    \Gamma_{f(Q), \xi} &=\left(
    \begin{array}{cccc}
    \{0,0,0,0\} & \{0,0,0,0\} & \{0,0,0,0\} & \{0,0,0,0\} \\
    \{0,0,0,0\} & \left\{0,\frac{B'}{B}-\xi\frac{B}{r}-\frac{1}{r},0,0\right\} & \left\{0,0,\xi \frac{  r}{B},0\right\} & \left\{0,0,0,\xi \frac{ r \sin ^2\theta}{B}\right\} \\
    \{0,0,0,0\} & \left\{0,0,-\xi \frac{B}{ r},0\right\} & \left\{0,-\xi \frac{B}{  r},0,0\right\} & \{0,0,0,-\sin\theta\cos\theta \} \\
    \{0,0,0,0\} & \left\{0,0,0,-\xi \frac{B}{r}\right\} & \{0,0,0,\cot\theta\} & \left\{0,-\xi \frac{B}{r},\cot\theta,0\right\} 
    \end{array}
    \right) \label{eq:fQ_xi},
\end{align}
where $\xi=\pm 1$. The corresponding $Q$ is
\begin{align}
&Q_{f(Q), \xi} =-\frac{2 (\xi  B+1) \left(2 r A'+\xi  A B+A\right)}{r^2 A B^2},\\
&Q_{f(Q), \xi, G\rightarrow 0}=-\frac{2(\xi+1)^2}{r^2}.
\end{align}
The St\"uckelberg fields are
\begin{align}
    \xi^a &= \{t, V(r) r\sin\theta\cos\phi, V(r)r\sin\theta\sin\phi, V(r)r\cos\theta\},\notag\\
    V(r) &= exp(\int \frac{-1-\xi B}{r}dr). 
\end{align}

Their EoMs are  
\begin{align}
    E_{f(Q), \xi,00}=&-\frac{A}{2 r^2 B^3}(\left(4 \xi  r A B^2 Q'+4 r A B Q'\right)f_{QQ}\notag\\
    &+(4 r B A' (\xi  B+1)-4 r A B'+4 \xi  A B^2+4 A B)f_Q+r^2 A B^3 f),\\
    E_{f(Q), \xi,11}=&\frac{1}{2r^2AB}( \left(4 r BA' (\xi  B+2)+4 \xi  A B^2+4 AB\right)f_Q+r^2 A B^3 f),\\
    E_{f(Q), \xi,22}=&\frac{E_{f(Q), \xi, 33}}{\sin^2\theta}=\frac{1}{2AB^3}((2 r^2 B A' Q'+2 r A B Q'+2 \xi  r A B^2 Q')f_{QQ}\notag\\
    &+ (2 r^2 B A''-2 r^2 A' B'+4 \xi  r B^2 A'+6 r B A'\notag\\
    &-2 r A B'+2 A B+4 \xi  AB^2 + 2 A B^3)f_Q+r^2 A B^3 f).
\end{align}

For $f(T)$ gravity, there are three branches (we call them $\xi$ branch and complex branch) in Weitzenb\"ock gauge. In order to facilitate comparison with $f(Q)$ case, the definition of $\xi$ in the tetrad field in this paper differs from that in \cite{Bahamonde:2021srr} by a minus sign: 
\begin{align}
    h_{f(T),\xi}{}^a{}_\mu&=\left(
        \begin{array}{cccc}
         A(r) & 0 & 0 & 0 \\
         0 & B(r) \sin\theta \cos\phi & -\xi r \cos \theta  \cos \phi  & \xi r \sin \theta  \sin \phi  \\
         0 & B(r) \sin \theta  \sin \phi  & -\xi r \cos \theta  \sin \phi  & -\xi r \sin \theta  \cos \phi  \\
         0 & B(r) \cos \theta  & \xi r \sin \theta  & 0 \\
        \end{array}
        \right), \xi = \pm 1,\label{tetrad1}\\
    h_c{}^a{}_\mu&=\left(
    \begin{array}{cccc}
    0 & i B(r) & 0 & 0 \\
    i A(r) \sin \theta  \cos \phi  & 0 & -r \sin \phi  & -r \sin \theta  \cos \theta  \cos \phi  \\
    i A(r) \sin \theta  \sin \phi  & 0 & r \cos \phi  & -r \sin \theta  \cos \theta  \sin \phi  \\
    i A(r) \cos \theta  & 0 & 0 & r \sin ^2\theta  \\
    \end{array}
    \right).\label{tetrad2}
\end{align}

Branch $\xi=-1$ has the same Lorentz vector in Weitzenb\"ock gauge as Eq.~\eqref{eq:Lorentz_vector_ME}. Branch $\xi=1$ and the complex branch have no Lorentz vector.

The corresponding torsion scalar is 
\begin{align}
    T_\xi&=-\frac{2 (\xi B+1) \left(2 r A'+A (1+\xi B\right))}{r^2 A B^2},\\
    T_c&=-\frac{2 \left(2 r A'+A \left(B^2+1\right)\right)}{r^2 A B^2}.
\end{align}
When gravity is switched off ($A(r)\rightarrow1,B(r)\rightarrow1$), they become
\begin{align}
    T_{\xi,G\rightarrow0}&=-\frac{2(1+\xi)^2}{r^2},\\
    T_{c,G\rightarrow0}&=-\frac{4}{r^2}.
\end{align} 
EoMs of $\xi$ branch are
\begin{align}
    E_{f(T), \xi,00}=&-\frac{A}{2r^2B^3}(\left(4 \xi r A B^2 T'+4  r A B T'\right)f_{TT}\notag\\
    &+\left(4 r B A' (\xi B+1 )f_T-4  r A B'+4\xi A B^2+4 A B\right)+  r^2 A B^3 f),\label{eq:fT_xi_EoM_0}\\
    E_{f(T), \xi,11}=&\frac{1}{2r^2AB}(\left(4 r BA' (\xi B+2 )f_T +4 \xi A B^2+4  AB\right)+r^2 A B^3 f),\label{eq:fT_xi_EoM_1}\\
    E_{f(T), \xi,22}=&\frac{E_{f(T), \xi,33}}{\sin^2\theta}=\notag\\
    &\frac{1}{2AB^3}((2 r^2 B A' T'+2 r A B T'+2 \xi  r A B^2 T')f_{TT}\notag\\
    &+ (2 r^2 B A''-2  r^2 A' B'+6  r B A'+4 \xi  r B^2 A'\notag\\
    &-2  r A B'+2 A B+4 \xi  A B^2+2 A B^3)f_T+ r^2 A B^3 f).\label{eq:fT_xi_EoM_2}
\end{align}

EoMs of the complex branch are
\begin{align}
    E_{c,00}=&-\frac{A}{2r^2B^3}(4 r A B T' f_{TT}\notag\\
    &+\left(4 r B A'-4 r A B'+4 AB\right) f_T+r^2 A B^3 f),\\
    E_{c,11}=&\frac{1}{2r^2AB}\left(4 B \left(2 r A'+A\right) f_T+r^2 A B^3 f\right),\\
    E_{c,22}=&\frac{E_{c,33}}{\sin^2\theta}=\frac{1}{2AB^3}( \left(2 r^2 B A' T'+2 r A B T'\right)f_{TT}\notag\\
    &+ (2 r^2 B A''-2 r^2 A' B'+6 r B A'-2 r AB'\notag\\
    &+2 AB+2 AB^3 )f_T+r^2 A B^3 f).
\end{align}

These results are summarized in Table \ref{tab:connection_branches_in_black_hole_spacetime} for convenience.

\begin{table*}
    \centering
    \caption{Different branches of f(Q) and f(T) theory in black hole background. $G\rightarrow0$ denotes the case when gravity vanishes, namely in Minkowski spacetime with $A(r)=1,B(r)=1$.}
    \vspace{10pt}
    \renewcommand{\arraystretch}{2}
    \begin{tabular}{cccccccc}
    \hline
    \multirow{4}{*}{$f(Q)$}& Branch & $Q$ & $Q_{G\rightarrow 0}$ &\multirow{4}{*}{$f(T)$}& Branch & $T$ & $T_{G\rightarrow 0}$\\
    \cmidrule(lr){2-4} \cmidrule(lr){6-8}
    &$\xi=1$ &$-\frac{2 (B+1) \left(2 r A'+A B+A\right)}{r^2 A B^2}$ &  $-\frac{8}{r^2}$  & &$\xi=1$  &$-\frac{2 (B+1) \left(2 r A'+A B+A\right)}{r^2 A B^2}$  & $-\frac{8}{r^2}$ \\
   \cmidrule(lr){2-4} \cmidrule(lr){6-8}
        &$\xi=-1$ 
    &$-\frac{2 (-B+1) \left(2 r A'-A B+A\right)}{r^2 A B^2}$& 0& &$\xi=-1$ &$-\frac{2 (-B+1) \left(2 r A'-A B+A\right)}{r^2 A B^2}$&0 \\
    \cmidrule(lr){2-4} \cmidrule(lr){6-8}
    & ME,Q
    & $\frac{2 \left(B^2-1\right) \left(B A'+A B'\right)}{r A B^3}$ & 0 && Complex &$-\frac{2 \left(2 r A'+A \left(B^2+1\right)\right)}{r^2 A B^2}$ & $-\frac{4}{r^2}$ \\
    \hline
    \end{tabular}
    \label{tab:connection_branches_in_black_hole_spacetime}
\end{table*}  

\subsection{Compare  connection branches between $f(Q)$ and $f(T)$ gravity}
\label{sec:compare_connection_branches}
Firstly, in any given spacetime, there exists at least one branch which turns out a vanishing geometrical trinity when gravity switches off to Minkowski spacetime. They are: 
\begin{enumerate}
    \item Branch \uppercase\expandafter{\romannumeral1} and Branch ME in cosmological spacetime (Table~\ref{tab:connection_branches_in_cosmological_spacetime}),
    \item Branch $\xi=-1$ and Branch ME in black hole spacetime (Table~\ref{tab:connection_branches_in_black_hole_spacetime}).
\end{enumerate}
Moreover, some of them in $f(Q)$ gravity have the same affine connection, which is independent of gravity. In Section~\ref{sec:Minkowski_eq_correspondence}, we call this relation as Minkowski-equivalence correspondence.

Secondly, we find for both cosmological spacetime and black hole spacetime, some branches have the same EoMs between $f(Q)$ and $f(T)$ gravity. They are:
\begin{enumerate}
    \item Branch \uppercase\expandafter{\romannumeral1} of $f(Q)$ gravity and Branch ME of $f(T)$ gravity in cosmological spacetime,
    \item $\xi$ Branch of $f(T)$ and $f(Q)$ gravity in black hole spacetime.
\end{enumerate}
In Section~\ref{sec:EoMs_Correspondence}, we call this correspondence as equations-of-motion (EoMs) correspondence. One question is whether there exists EoMs correspondence for the complex branch Eq.~\eqref{tetrad2} of $f(T)$ in the black hole spacetime. If it does, we can conclude solutions of $f(T)$ in the black spacetime are just a subset of solutions of $f(Q)$ gravity. However, using the general expression of affine connection, we find this correspondence doesn't exist.

\section{Minkowski-equivalence correspondence between $f(Q)$ and $f(T)$ gravity}
\label{sec:Minkowski_eq_correspondence}

\subsection{General spin connection in $f(Q)$ gravity }

As we discussed in  Section~\ref{sec:definitions},   the
metric-affine formulation and the spin-tetrad formulation are two equivalent 
descriptions of the same physical system. 
Due to the different advantages they offer for solving geometrical constraints, 
we select the different preferred formulations: the metric-affine for GR and 
the tetrad-spin for TG. While our initial intuition in Symmetric Teleparallel 
Gravity might lead us to favor the metric-affine approach, due to its 
torsionless condition, it becomes necessary to adopt the tetrad-spin formulation 
to facilitate comparisons between different branches of $f(T)$ and $f(Q)$ 
gravity. This choice is particularly relevant since the complex branch in $f(T)$ 
emerges from the tetrad-spin formulation.

Firstly, the flat condition constrains the spin connection to the form 
\begin{equation}
    A^a{}_{b\mu}=(N^{-1})^a{}_c \partial_\mu N^c{}_b,
\end{equation}
where $N^a{}_b$ are components of a matrix belonging to the general linear group $GL(4, \mathbb{R})$. 
In order to implement the torsionless condition, instead of solving Eq.~\eqref{eq: curvature tetrad-spin} directly, we utilize Eq.~\eqref{eq: spin 
connection definition} and Eq.~\eqref{eq: coincident affine connection} to 
derive
\begin{align}
    A^a{}_{b \mu}&=h^a{}_\rho(\partial_\mu h_b{}^\rho+\Gamma^\rho{}_{\nu \mu}h_b{}^\nu) \notag\\
    &= \frac{\partial x^\rho}{\partial\xi^\alpha}h^a{}_\rho \partial_\mu(\frac{\partial \xi^\alpha}{\partial x^\nu}h_b{}^\nu), \label{eq:spin_general_expression}
\end{align}
which allows us to deduce the form of $N^a{}_b$ as:
\begin{align}
    N^a{}_b\equiv\delta^a{}_\alpha \frac{\partial \xi^\alpha}{\partial x^\nu}h_b{}^\nu.
\end{align}

\subsection{Tetrad-spin formulation of $f(Q)$ gravity \label{sec:tetrad-spin_fQ}}

The action of $f(Q)$ gravity in tetrad-spin formulation is defined as: 
\begin{align}
        \mathcal{S}_{f(Q)}=-\frac{1}{2 \kappa}\int d^4x \, h f(Q)+\mathcal{S}_{matter}.
\end{align}
As is well known, the tetrad  and spin connection are two independent variables.
To derive the field equations with respect to these variables  we perform   
variation of the action using   Eq.~\eqref{eq: non-metricity tetrad-spin}. The 
resulting field equations for the tetrad are given by:
\begin{align}
  &\frac{1}{2}f_Q[-2{h_a}^\rho Q_{\alpha \rho \nu}P^{\alpha \nu \mu}+{h_a}^\alpha g^{\beta \mu}(P_{(\beta|\nu \rho}{Q_{\alpha)}}^{\nu \rho}+2P^\nu{}_{\rho(\alpha}Q_{|\nu|}{}^{\rho}{}_{\beta)})]+\frac{1}{2}f\,{h_a}^\mu
  =\kappa {\mathcal{\tilde{T}}_a}^\mu,
\end{align}
where ${\mathcal{\tilde{T}}_a}^\mu \equiv \frac{1}{h}\frac{\delta 
\mathcal{L}_{matter}}{\delta {h^a}_\mu}(h^a{}_\mu, A^a{}_{b\mu})$. 
Additionally, variation of the action with respect to the spin connection 
leads to
\begin{align}
  \delta_A \mathcal{S}_{f(Q)}=-\frac{1}{2\kappa}\int d^4x \,2hf_Q P^\mu{}_a{}^b \delta{A^a}_{b \mu} +\delta_A\mathcal{S}_{matter}.
\end{align}
However, this approach reveals that the variation of the action with  respect to 
the tetrad does not yield the same field equations with Eq.~\eqref{eq: f(Q) 
metric EoM}, as it lacks the necessary dynamical degrees of freedom, indicating 
that this function acts merely as a constraint. To address this issue, we can 
use the torsionless and flat conditions to eliminate the spin connection, as 
represented in Eq.~\eqref{eq:spin_general_expression}. The variation of the spin 
connection can be decomposed into the variation of the tetrad and the variation 
of the St\"uckelberg fields,
\begin{align}
  \delta A^a{}_{b \mu} = \delta_h A^a{}_{b \mu} + \delta_\xi A^a{}_{b \mu}.
\end{align}
This leads us to the more reasonable field equations for the tetrad, namely
\begin{align}
  &\frac{h_a{}^\rho}{h}\nabla_\nu(hf_QP^\nu{}_\rho{}^\mu)+\frac{1}{2}f_Q[-2{h_a}^\rho Q_{\alpha \rho \nu}P^{\alpha \nu \mu}+{h_a}^\alpha g^{\beta \mu}(P_{(\beta|\nu \rho}{Q_{\alpha)}}^{\nu \rho}+2P^\nu{}_{\rho(\alpha}Q_{|\nu|}{}^{\rho}{}_{\beta)})]\notag \\
  &+\frac{1}{2}f\,{h_a}^\mu
  =\kappa {\mathcal{T}_a}^\mu, \label{eq: field equation, tetrad}
\end{align}
where 
\begin{align}
{\mathcal{T}_a}^\mu \equiv \frac{1}{h}\frac{\delta \mathcal{L}_{matter}}{\delta {h^a}_\mu}(h^a{}_\mu, \xi^\alpha)={\mathcal{\tilde{T}}_a}^\mu+\frac{1}{h}\frac{\delta \mathcal{L}_{matter}}{\delta A^c{}_{b\nu}}\frac{\delta A^c{}_{b\nu}}{\delta h^a{}_\mu}.
\end{align}

To derive the field equations for the St\"uckelberg fields, we calculate the 
variation of the spin connection directly  and obtain the identity:
\begin{align}
  \delta_\xi A^a{}_{b \mu} = h^a{}_\rho h_b{}^\nu \frac{\partial x^\rho}{\partial \xi^\alpha}\nabla_\mu\nabla_\nu \delta \xi^\alpha.\label{eq: variation of spin connection}
\end{align}
Utilizing Eq.~\eqref{eq: variation of spin connection} as well as Eq.~\eqref{eq: 
affine connection definition}, we  can derive 
\begin{align}
    \delta_\xi \Gamma^\alpha{}_{\mu \nu} = \frac{\partial x^\alpha}{\partial  
\xi^\lambda}\nabla_\nu \nabla_\mu \delta \xi^\lambda. \label{eq: variation of 
affine connection}
\end{align}
Moreover, using Eq.~\eqref{eq: variation of spin connection} and Eq.~\eqref{eq: 
variation of affine connection}, we can express the variation of the action with 
respect to the St\"uckelberg fields as:
\begin{align}
  \delta_\xi \mathcal{S}_{f(Q)}=&-\frac{1}{2\kappa}\int d^4x \, \nabla_\nu 
\nabla_\mu(2hf_Q{{P^\mu}_\rho}^\nu\frac{\partial x^\rho}{\partial \xi^\alpha}) 
\delta \xi^\alpha+\frac{1}{2\kappa}\int d^4x \,\nabla_\mu\nabla_\nu 
(\mathcal{H}_\alpha{}^{\mu \nu}\frac{\partial x^\alpha}{\partial 
\xi^\lambda})\delta\xi^\lambda.
\end{align}
Employing the identity $
  \nabla_\mu \frac{\partial x^\rho}{\partial \xi^\alpha}\equiv 0$, which can be 
proved by Eq.~\eqref{eq: coincident affine connection}, we obtain
\begin{align}
    \delta_\xi \mathcal{S}_{f(Q)}=&-\frac{1}{2\kappa}\int d^4x  \,\frac{\partial 
x^\rho}{\partial \xi^\alpha}[\nabla_\nu \nabla_\mu(2hf_Q{{P^\mu}_\rho}^\nu) -\nabla_\mu\nabla_\nu (\mathcal{H}_\rho{}^{\mu \nu})]\delta \xi^\alpha, 
\end{align}
and thus the field equations of St\"uckelberg fields are extracted as
\begin{align}
    \frac{\partial x^\rho}{\partial \xi^\alpha}[2\nabla_\nu  \nabla_\mu(hf_Q 
P^{\mu \nu}{}_\rho)-\nabla_\mu \nabla_\nu\mathcal{H}_\rho{}^{\mu \nu}]=0. 
\label{eq: Stuckelberg field equation}
\end{align}
As we see, it differs from Eq.~\eqref{eq: f(Q) affine EoM} by a 
factor of $\frac{\partial x^\rho}{\partial \xi^\alpha}$, and thus in principle 
it possesses a broader range of solutions. This is due to the fact that even 
after fixing the affine connection, there remain residual degrees of freedom in 
the St\"uckelberg fields. From Eq.~\eqref{eq: coincident affine connection}, we 
observe that under the transformation $\xi^\alpha \rightarrow M^\alpha{}_\beta 
\xi^\beta$, where $M^\alpha{}_\beta$ is a coordinate-independent constant 
matrix, the affine connection remains invariant. With such a transformation, the 
terms inside the square bracket of Eq.~\eqref{eq: Stuckelberg field equation} 
are unaffected, while $\frac{\partial x^\rho}{\partial \xi^\alpha}$ can acquire
an arbitrary value, leading to 
\begin{align}
    2\nabla_\nu \nabla_\mu(hf_Q P^{\mu \nu}{}_\rho)=\nabla_\mu \nabla_\nu\mathcal{H}_\rho{}^{\mu \nu}. \label{eq: field equation, affine connection}
\end{align}
As we observe, both Eq.~\eqref{eq: field equation, tetrad} and Eq.~\eqref{eq: 
field equation, affine connection} are identical to Eq.~\eqref{eq: f(Q) metric 
EoM} and Eq.~\eqref{eq: f(Q) affine EoM} respectively, which originate from the 
metric-affine formulation. Our approach indicates that the true equations of motions of $f(Q)$ gravity come from the variation of tetrad and St\"uckelberg fields, rather than the spin connection. This is easy to understand if we assume the Weitzenb\"ock definition of teleparallel gravity is the fundamental one and the procedure of St\"uckelberg formulation is a way to recover the covariance of theory \cite{Krssak:2024xeh}.

In our current work, to preserve the generality of our conclusions, we do not 
assume a vanishing hypermomentum tensor; instead, we allow it to be determined 
by the affine field equations. Therefore, in the following section, we will 
focus solely on presenting the metric field equations.
The detailed calculations of these field equations and the  proofs of the 
identities are provided in Appendix \ref{appendix: EoMs}.

\subsection{Minkowski-equivalence correspondence}

\textit{Definition: \\
For every branch of $f(T)$ gravity with a vanishing torsion tensor when gravity is switched off, if there exists a corresponding branch in $f(Q)$ gravity which has St\"uckelberg fields with the same components as the Lorentz vector of $f(T)$ gravity in the Weitzenb\"ock gauge, we call this correspondence as \textbf{Minkowski-equivalence correspondence}.}

To demonstrate the existence and practical utility of this correspondence, we begin with Eq.~\eqref{eq:spin_general_expression}. Eq.~\eqref{eq:spin_general_expression} tells us once we have the St\"uckelberg fields and tetrad, the spin connection is determined. The key question is how to find the St\"uckelberg fields. Minkowski-equivalence correspondence provides us with a new way to solve this problem.

If we assume in $f(Q)$ gravity: 
\begin{enumerate}
    \item a vanishing non-metricity tensor when gravity is switched off,
    \item the affine connection is independent of gravity,
\end{enumerate}
we 
can further simplify Eq.~\eqref{eq:spin_general_expression} \textbf{as a function of tetrad only}. In Minkowski spacetime, where 
non-metricity tensor is zero, the tetrad takes the same form as in Eq.~\eqref{eq:tetrad_Lorentz_vector} in the case of TG. By imposing the Weitzenb\"ock 
gauge with a vanishing Lorentz connection after applying a Lorentz 
transformation $\Lambda^a{}_b$, we can express the tetrad in the form
\begin{align}
    h_{(r)}{}^a{}_\mu = \Lambda^a{}_b \tilde{h}_{(r)}{}^b{}_\mu= 
\Lambda^a{}_b\partial_\mu \tilde{v}^b ,
    \label{eq: Lorentz vector} 
\end{align}
where $r$ denotes quantities in Minkowski spacetime and $\tilde{v}^a$ is the Lorentz vector in the Weitzenb\"ock gauge.

On the other hand, if the affine connection is independent of gravity, when gravity is absent,  there exists a 
global coordinate transformation that satisfies
\begin{equation}
    g_{\mu\nu}=\frac{\partial \xi^\alpha}{\partial x^\mu}\frac{\partial 
\xi^\beta}{\partial x^\nu}\eta_{\alpha\beta}, 
\end{equation}
where $\eta_{\alpha \beta}=diag\{-1,1,1,1\}$ and $\xi^\alpha$ is the St\"uckelberg fields. Therefore, if we define $\xi^a 
\equiv \delta^a{}_\alpha\xi^\alpha$, implying $\xi^a$ and $\xi^\alpha$ have the 
same components despite differing in the index type, the tetrad can be 
expressed as 
\begin{equation}
    {h^a}_\mu=\frac{\partial \xi^a}{\partial x^\mu}.
\end{equation}
This equation  indicates simply that the $\xi^\alpha$ has the same components as 
the Lorentz vector in the Weitzenb\"ock gauge, leading to the conclusion 
$\xi^\alpha = \delta_a{}^\alpha v^a$. 

This straightforward conclusion is useful because the Lorentz vector is determined by the tetrad only then the spin connection of $f(Q)$ gravity has one solution (branch) that is determined by the tetrad only.

Consequently, we obtain a simplified formula of the spin connection, namely
\begin{align}
    A^a{}_{b\mu}&=(N^{-1})^a{}_c \partial_\mu N^c{}_b, \label{eq: spin special 1}\\
    N^a{}_b &= \tilde{h}_{(r)}{}^a{}_\mu h_b{}^\mu. \label{eq: spin special 2}
\end{align}
By choosing the spin connection of $f(Q)$ gravity in the form given by Eq. 
\eqref{eq: spin special 2}, we establish a correspondence between $f(Q)$ and $f(T)$ gravity. 

In summary, the Minkowski-equivalence tetrad-spin formulation of STG can be 
explicitly articulated through the following steps:
\begin{enumerate}
    \item Choose one arbitrary tetrad.
    \item Switch off   gravity by removing parameters containing 
gravitational information, in order  to obtain the tetrad in Minkowski 
spacetime.
    \item Apply a Lorentz transformation to achieve the tetrad in Weitzenb\"ock gauge.
    \item Use Eq. \eqref{eq: spin special 1} and Eq. \eqref{eq: spin special 2} to calculate the spin connection.  
\end{enumerate}


With this correspondence, let's see the first finding in Section~\ref{sec:compare_connection_branches} from a new perspective.

In spherical coordinates, tetrads in black hole spacetime Eq.~\eqref{eq:tetrad_blackhole} and cosmological spacetime Eq.~\eqref{eq:tetrad_cosmological} degenerate into the same Minkowski spacetime tetrad: 
\begin{equation}
    h_{(r)}{}^a{}_\mu = diag\{1, 1, r, r\sin\theta\}, \label{eq: Minkowski tetrad}
\end{equation}
with the non-vanishing components of the Lorentz connection given by:
\begin{align}
    {\omega^r}_{\theta\theta} &= -1,\quad &{\omega^r}_{\phi\phi} &= -\sin\theta, \notag\\
    {\omega^\theta}_{r\theta} &= 1, \quad &{\omega^\theta}_{\phi\phi}&=-\cos\theta, \notag\\
    {\omega^\phi}_{r\phi} &= \sin\theta, \quad &{\omega^\phi}_{\theta\phi}&= \cos\theta.
\end{align}
To restore the Weitzenb\"ock gauge, we apply the following Lorentz transformation:
\begin{equation}
    {\Lambda^a}_b =
    \left(
    \begin{array}{cccc}
        1 & 0 & 0 & 0 \\
        0 & \sin\theta\cos\phi & \cos\theta\cos\phi & -\sin\phi \\
        0 & \sin\theta\sin\phi & \cos\theta\sin\phi & \cos\phi \\
        0 & \cos\theta & -\sin\theta & 0
    \end{array}
    \right).
\end{equation}
In this new tangent coordinate system, the Lorentz connection vanishes and the  
tetrad becomes
\begin{equation}
    \tilde{h}_{(r)}{}^a{}_\mu =
    \left(
    \begin{array}{cccc}
        1 & 0 & 0 & 0 \\
        0 & \sin\theta\cos\phi & r\cos\theta\cos\phi & -r\sin\theta\sin\phi \\
        0 & \sin\theta\sin\phi & r\cos\theta\sin\phi & r\sin\theta\cos\phi \\
        0 & \cos\theta & -r\sin\theta & 0
    \end{array} 
    \right). \label{eq:tetrad_ME}
\end{equation}
Using Eq. \eqref{eq: Lorentz vector}, the Lorentz vector is
\begin{equation}
    v^a = \{t, r\sin\theta\cos\phi, r\sin\theta\sin\phi, r\cos\theta\}.
\end{equation}
This expression corresponds to the coordinate transformation from spherical to 
Cartesian coordinates.  Since $\xi^\alpha = \delta_a{}^\alpha v^a$, we obtain the affine connection of STG in 
spherical coordinates as
\begin{align}
    {\Gamma^r}_{\theta\theta} &= -r, \quad &{\Gamma^r}_{\phi\phi} &= -r\sin^2\theta, \notag\\
    {\Gamma^\theta}_{r\theta} &= {\Gamma^\theta}_{\theta r} =\frac{1}{r}, \quad 
&{\Gamma^\theta}_{\phi\phi}&=-\cos\theta \sin\theta,\notag\\ 
    {\Gamma^\phi}_{r\phi} &= {\Gamma^\phi}_{\phi r}=\frac{1}{r}, \quad 
&{\Gamma^\phi}_{\theta\phi} &= {\Gamma^\phi}_{\phi \theta}=\cot\theta.  
\label{eq:connecton_from_ME}
\end{align}
Additionally, the non-vanishing components of the corresponding spin connection 
with respect to Eq. \eqref{eq:tetrad_blackhole} are
\begin{align}
    A^t{}_{tr}&=-\frac{A'}{A}, \quad & A^r{}_{rr}&=-\frac{B'}{B}, \notag\\
    A^r{}_{\theta \theta}&= \frac{A^r{}_{\phi \phi}}{\sin\theta}=-B,  \quad & 
A^\theta{}_{r\theta}&=\frac{A^\phi{}_{r\phi}}{\sin\theta}=\frac{1}{B}, 
\notag\\
    A^\theta{}_{\phi \phi}&=-A^\phi{}_{\theta \phi}=-\cos\theta, 
\end{align}
while the non-vanishing components of the corresponding spin connection with 
respect to Eq. \eqref{eq:tetrad_cosmological} are
\begin{align}
    A^r{}_{r t}&=A^\theta{}_{\theta t}=A^\phi{}_{\phi t}=- \frac{a'}{a}, \quad & 
A^r{}_{\theta \theta}&=-A^\theta{}_{r\theta}=-1,\notag\\
    A^r{}_{\phi \phi}&=-A^\phi{}_{r\phi}=-\sin\theta,  \quad & 
A^\theta{}_{\phi \phi }&=-A^\phi{}_{\theta \phi}=-\cos\theta.
\end{align} 
As we observe, these two spin connections are no longer antisymmetric in their 
first two indices and now include metric components, which give rise to 
dynamical effects in the spin connection within $f(Q)$ gravity.

If the spin connection in the ME branch of $f(T)$ gravity is interpreted as an inertial effect, then all gravitational effects arise solely from the tetrad field. In contrast, in the ME branch of $f(Q)$ gravity, the gravitational contributions from the spin connection and the tetrad field cancel each other out, thereby restoring the trivial affine connection of Minkowski spacetime.

Eq.~\eqref{eq:connecton_from_ME} is the same as $\Gamma_{ME,Q}$~\eqref{eq:min_affine} and Branch \uppercase\expandafter{\romannumeral1} with $\gamma=0$ of $f(Q)$ gravity. Branch ME~\eqref{eq:spin_ME} and Branch $\xi=-1$ of $f(T)$ gravity just have the same tetrad as Eq.~\eqref{eq:tetrad_ME}. They both lead to a vanishing geometrical trinity when gravity switches off. These are what we find in Section~\ref{sec:compare_connection_branches}.

\section{Equations-of-motion correspondence between $f(Q)$ and $f(T)$ gravity 
\label{sec:EoMs_Correspondence}}

The Minkowski-equivalent approach is useful for that it establishes a 
bijective mapping between some of the  solutions of $f(Q)$ and $f(T)$ gravity. 
However, an additional equivalence exists even in non-vanishing gravity 
scenarios. In \cite{DAmbrosio:2021zpm}, the authors used a symmetry method to 
constrain the form of affine connection both in $f(Q)$ and $f(T)$ gravity. In 
particular, they found two cases in which the field 
equations for $f(Q)$ and $f(T)$ gravity have identified forms, producing the 
same solutions. Hence, we call this correspondence  "equations-of-motion 
(EoMs) correspondence". For more transparency, we 
prompt another practical approach to establish  this correspondence between 
$f(Q)$ and $f(T)$ gravity.

In order to find the corresponding affine connection between $f(Q)$ and $f(T)$ 
gravity,  there are two conditions that should be satisfied:
\begin{enumerate}
    \item The non-metricity scalar in $f(Q)$ gravity should have the same value 
with the torsion scalar in $f(T)$ gravity  at the same 
spacetime point, namely \label{cond1}
        \begin{align}
            Q_{f(Q)} = T_{f(T)}. \label{eq:cond1}
        \end{align}
    \item The field equations in $f(Q)$ gravity should take the same form as those in $f(T)$ gravity
(regardless of the functional forms of $T$, $Q$ and $f$), namely \label{cond2}
        \begin{align}
            E_{\mu \nu, f(Q)} = E_{\mu \nu, f(T)}. \label{eq:cond2}
        \end{align} 

\end{enumerate}

Below we will analyze in detail the process of deriving the EoMs correspondence 
between $f(Q)$ and $f(T)$ gravity in the static and spherically symmetric 
spacetime. Furthermore, we will   briefly discuss the correspondence in the 
cosmological spacetime.

\subsection{General affine connection of $f(Q)$ gravity in static and 
spherically  symmetric spacetime}

Some research point out there are two general branches which are able to produce beyond-GR solutions \cite{DAmbrosio:2021zpm,Heisenberg:2023lru}. The first one (we call it General A) is:
\begin{align}
    \Gamma_{General \, A} = \left(
\begin{array}{cccc}
 \{0,0,0,0\} & \left\{0,-\frac{m}{\Gamma^r{}_{\theta \theta}(r)^2},0,0\right\} & \{0,0,m,0\} & \left\{0,0,0,m \sin ^2\theta \right\} \\
 \{0,0,0,0\} & \left\{0,-\frac{\Gamma^r{}_{\theta \theta}'(r)+1}{\Gamma^r{}_{\theta \theta}(r)},0,0\right\} & \{0,0,\Gamma^r{}_{\theta \theta}(r),0\} & \left\{0,0,0,\sin ^2\theta \, \Gamma^r{}_{\theta \theta}(r)\right\} \\
 \{0,0,0,0\} & \left\{0,0,-\frac{1}{\Gamma^r{}_{\theta \theta}(r)},0\right\} & \left\{0,-\frac{1}{\Gamma^r{}_{\theta \theta}(r)},0,0\right\} & \{0,0,0,-\sin \theta \cos \theta \} \\
 \{0,0,0,0\} & \left\{0,0,0,-\frac{1}{\Gamma^r{}_{\theta \theta}(r)}\right\} & \{0,0,0,\cot \theta \} & \left\{0,-\frac{1}{\Gamma^r{}_{\theta \theta}(r)},\cot \theta ,0\right\} \\
\end{array}
\right),
\label{eq:f(Q)_general_connection_A}
\end{align}
where $m$ is an arbitrary constant and $\Gamma^r{}_{\theta \theta}(r)$ is an arbitrary function determined by the symmetric components of the metric field equations.

The second one (we call it General B) is:
\begin{align}
    \Gamma^t{}_{\mu\nu}&=\left(
    \begin{array}{cccc}
     \frac{k}{2}-c & \frac{k \left(\frac{k}{2 (2 c-k)}+1\right)}{2 c \Gamma^r{}_{\theta\theta}} & 0 & 0 \\
     \frac{k \left(\frac{k}{2 (2 c-k)}+1\right)}{2 c \Gamma^r{}_{\theta\theta}} & \frac{k \left(8 c^2+2 c k-k^2\right)}{8 c^2 (2 c-k)^2 \Gamma^r{}_{\theta\theta}^2} & 0 & 0 \\
     0 & 0 & \frac{k}{2 c (2 c-k)} & 0 \\
     0 & 0 & 0 & \frac{k \sin ^2\theta}{2 c (2 c-k)} \\
    \end{array}
    \right),
    \notag\\
    \Gamma^r{}_{\mu\nu}&=\left(
    \begin{array}{cccc}
     -c (2 c-k) \Gamma^r{}_{\theta\theta} & c+\frac{k}{2} & 0 & 0 \\
     c+\frac{k}{2} & -\frac{\frac{8 c^2+k^2}{8 c^2-4 c k}+\Gamma^r{}_{\theta \theta}'}{\Gamma^r{}_{\theta\theta}} & 0 & 0 \\
     0 & 0 & \Gamma^r{}_{\theta\theta} & 0 \\
     0 & 0 & 0 & \sin ^2\theta \Gamma^r{}_{\theta\theta} \\
    \end{array}
    \right),
    \notag\\
    \Gamma^\theta{}_{\mu\nu}&=\left(
    \begin{array}{cccc}
     0 & 0 & c & 0 \\
     0 & 0 & -\frac{\frac{k}{2 (2 c-k)}+1}{\Gamma^r{}_{\theta\theta}} & 0 \\
     c & -\frac{\frac{k}{2 (2 c-k)}+1}{\Gamma^r{}_{\theta\theta}} & 0 & 0 \\
     0 & 0 & 0 & -\sin \theta  \cos \theta  \\
    \end{array}
    \right),
    \Gamma^\phi{}_{\mu\nu}=\left(
    \begin{array}{cccc}
     0 & 0 & 0 & c \\
     0 & 0 & 0 & -\frac{\frac{k}{2 (2 c-k)}+1}{\Gamma^r{}_{\theta\theta}} \\
     0 & 0 & 0 & \cot \theta  \\
     c & -\frac{\frac{k}{2 (2 c-k)}+1}{\Gamma^r{}_{\theta\theta}} & \cot \theta & 0 \\
    \end{array}
    \right),
    \label{eq:f(Q)_general_connection_B}
\end{align}
where $c$ and $k$ are arbitray constants ($c \ne 0,k \ne 2c$) and $\Gamma^r{}_{\theta \theta}(r)$ is an arbitrary function determined by the symmetric components of the metric field equations.

Different from the method used in \cite{DAmbrosio:2021zpm}, here we adopt the metric-affine theory to derive the general form of the affine connection in flat, torsion-free, static and spherically symmetric spacetime: 
\begin{align}
    \Gamma=
\left(
\begin{array}{cccc}
 \{0,0,0,0\} &  \left\{0,\frac{C_2'}{k1}-\frac{C_2 C_5''}{k_1C_5'},0,0\right\} & 
\left\{0,0,\frac{C_2 C_5}{k_1 C_5'},0\right\} & \left\{0,0,0,\frac{C_2 C_5 \sin 
^2\theta}{k_1 C_5'}\right\} \\
 \{0,0,0,0\} & \left\{0,\frac{C_5''}{C_5'},0,0\right\} &  
\left\{0,0,-\frac{C_5}{C_5'},0\right\} & \left\{0,0,0,-\frac{C_5 \sin ^2\theta 
}{C_5'}\right\} \\
 \{0,0,0,0\} & \left\{0,0,\frac{C_5'}{C_5},0\right\} &  
\left\{0,\frac{C_5'}{C_5},0,0\right\} & \{0,0,0,-\sin \theta\cos\theta\} 
\\
 \{0,0,0,0\} &  \left\{0,0,0,\frac{C_5'}{C_5}\right\} & \{0,0,0,\cot \theta \} 
& \left\{0,\frac{C_5'}{C_5},\cot \theta,0\right\} \\
\end{array}
\right),
\label{eq: affine connection in SS}
\end{align}
where $C_2(r)$, $C_5(r)$ are functions of $r$ and $k_1$ is a constant. The 
derivation of this affine connection is  presented in Appendix 
\ref{appendix:metric_afffine_to_general_A}.
Defining    
\begin{align}
    \Gamma^r{}_{\theta\theta}(r)&\equiv-\frac{C_5}{C_5'},\\
    m(r)&\equiv \frac{C_2 C_5}{k_1 C_5'},
\end{align}
the above form can be simplified to
\begin{align}
    \Gamma = \left(
\begin{array}{cccc}
 \{0,0,0,0\} &  \left\{0,-\frac{\Gamma^r{}_{\theta\theta} 
m'+m}{\Gamma^r{}_{\theta\theta}{}^2},0,0\right\} & \{0,0,m,0\} & 
\left\{0,0,0,\sin ^2\theta\, m\right\} \\
 \{0,0,0,0\} &  
\left\{0,-\frac{\Gamma^r{}_{\theta\theta}'+1}{\Gamma^r{}_{\theta\theta}},0,
0\right\} & \{0,0,\Gamma^r{}_{\theta\theta},0\} & \left\{0,0,0,\sin ^2\theta  
\Gamma^r{}_{\theta\theta}\right\} \\
 \{0,0,0,0\} &  \left\{0,0,-\frac{1}{\Gamma^r{}_{\theta\theta}},0\right\} & 
\left\{0,-\frac{1}{\Gamma^r{}_{\theta\theta}},0,0\right\} & \{0,0,0,-\sin 
\theta\cos\theta \} \\
 \{0,0,0,0\} &  \left\{0,0,0,-\frac{1}{\Gamma^r{}_{\theta\theta}}\right\} & 
\{0,0,0,\cot \theta\} & \left\{0,-\frac{1}{\Gamma^r{}_{\theta\theta}},\cot 
\theta ,0\right\} \\
\end{array}
\right). \label{eq: affine connection in SS simplified}
\end{align}

This form has the same equations of motion as Eq.~\eqref{eq:f(Q)_general_connection_A}. In \cite{DAmbrosio:2021zpm}, they solved the off-diagonal components of the field equations to derive Eq.~\eqref{eq:f(Q)_general_connection_A}. However, in our formalism, when the affine connection is expressed in the form 
given by Eq.~\eqref{eq: affine connection in SS}, the off-diagonal components of 
the field equations vanish automatically. The diagonal components of the field 
equations are presented in Appendix \ref{appendix:EoMs_metric_affine_to_general_A}.

Since Eq.~\eqref{eq:general_affine_con_f(Q)_ss} is one of the parameterizations of flat connection, we can't guarantee Eq.~\eqref{eq: affine connection in SS} is the most general one. That's why General B~\eqref{eq:f(Q)_general_connection_B} can't be included in our general affine connection.

\subsection{Equations-of-motion correspondence in static and 
spherically symmetric spacetime}

To derive the correspondences based on the equations of motion, we first apply Condition~\hyperref[cond1]{2}. In this case, the field equations of $f(Q)$ and $f(T)$ 
gravity can be expressed as  
\begin{align}
    E_{\mu \nu,f(Q)}&=\kappa_0f+\kappa_1f_Q+\kappa_2Q'f_{QQ},\\
    E_{\mu \nu,f(T)}&=\kappa_0f+\kappa_1f_T+\kappa_2T'f_{TT}.
\end{align}
Proceeding forward, we use the equation
\begin{align}
    \frac{\kappa_2}{\kappa_0}|_{f(Q)}&=\frac{\kappa_2}{\kappa_0}|_{f(T)} \label{eq:EoM_correspondence_eq}
\end{align}
to determine parameters in Eq. \eqref{eq:f(Q)_general_connection_A} and Eq. \eqref{eq:f(Q)_general_connection_B}.

EoMs of General A~\eqref{eq:f(Q)_general_connection_A} and General B~\eqref{eq:f(Q)_general_connection_B} are presented in Appendix~\ref{appendix:EoMs_general_affine_connection}. Using these EoMs, we can solve for $\Gamma^r{}_{\theta\theta}$ according to Eq.~\eqref{eq:EoM_correspondence_eq}. In Table~\ref{tab:solutions_EoM_coorepondence}, we present these solutions for the two general branches of $f(Q)$ gravity and three tetrads of $f(T)$ gravity. 

For the $\xi$ branch in $f(T)$ gravity~\eqref{tetrad1} and the General A branch~\eqref{eq:f(Q)_general_connection_A} in $f(Q)$ gravity, we calculate 
\begin{alignat}{3}
    E_{00} &\rightarrow \frac{\kappa_2}{\kappa_0}|_{f(T)} = \frac{4 (\xi B+1 )}{  r B^2}, \frac{\kappa_2}{\kappa_0}|_{f(Q)}=-\frac{A^2 \left(B^2 (\Gamma^r{}_{\theta \theta})^2+r^2+2 r \Gamma^r{}_{\theta \theta}\right)}{r^2 B^2 \Gamma^r{}_{\theta \theta} }  &\rightarrow \Gamma^r{}_{\theta\theta} &= \xi \frac{r}{B}, \\
    E_{11} &\rightarrow \frac{\kappa_2}{\kappa_0}|_{f(T)} = 0,  \frac{\kappa_2}{\kappa_0}|_{f(Q)}=\frac{2 \left(\frac{B^2 \Gamma^r{}_{\theta \theta} }{r^2}-\frac{1}{\Gamma^r{}_{\theta \theta} }\right)}{B^2}&\rightarrow \Gamma^r{}_{\theta\theta} &= \pm \frac{r}{B}, \\
    E_{22} &\rightarrow \frac{\kappa_2}{\kappa_0}|_{f(T)} = \frac{2 \left(\xi  r A'+A (B+\xi )\right)}{\xi  r A B^2},  \frac{\kappa_2}{\kappa_0}|_{f(Q)}=\frac{2 \left(r \Gamma^r{}_{\theta \theta}  A'+A (\Gamma^r{}_{\theta \theta} +r)\right)}{r A B^2 \Gamma^r{}_{\theta \theta} }&\rightarrow \Gamma^r{}_{\theta\theta} &= \xi \frac{r}{B},
\end{alignat}
so solutions are
\begin{align}
    \Gamma^r{}_{\theta\theta}=\xi \frac{r}{B}.
\end{align}
We can check this will lead to the same field equations as Eqs.~\eqref{eq:fT_xi_EoM_0}--\eqref{eq:fT_xi_EoM_2} in $f(T)$ gravity.

For the complex solutions in $f(T)$ gravity, Eq.~\eqref{tetrad2}, solutions of General A and B are
\begin{align}
&\text{General A:} \quad
\left\{
\begin{aligned}
    E_{00} &\rightarrow \Gamma^r{}_{\theta\theta} = \pm i\frac{r}{B}, \\
    E_{11} &\rightarrow \Gamma^r{}_{\theta\theta} = \pm \frac{r}{B}, \\
    E_{22} &\rightarrow \Gamma^r{}_{\theta\theta} = \pm\infty,
\end{aligned}
\right. \label{eq:sol_generalA_complex}\\
&\text{General B:} \quad
\left\{
\begin{aligned}
    E_{00} &\rightarrow \Gamma^r{}_{\theta\theta} = \pm \frac{i(4c - k)}{\sqrt{4c (2c - k) (2A^2 - 2c^2 r^2 + c k r^2)}} \frac{Ar}{B}, \\
    E_{11} &\rightarrow \Gamma^r{}_{\theta\theta} = \pm \frac{(4c - k)}{\sqrt{4c (2c - k) (2A^2 + 2c^2 r^2 - c k r^2)}} \frac{Ar}{B}, \\
    E_{22} &\rightarrow \Gamma^r{}_{\theta\theta} = \pm \frac{i(4c - k)}{2c (2c - k)} \frac{A}{B}.
\end{aligned}
\right.\label{eq:sol_generalB_complex}
\end{align}
Eq.~\eqref{eq:sol_generalA_complex} means there is no correspondence in the general A branch of $f(Q)$ gravity for the complex solution in $f(T)$ gravity. For the General B branch \eqref{eq:f(Q)_general_connection_B} in $f(Q)$ gravity, solutions satisfying Eq. \eqref{eq:sol_generalB_complex} are 
\begin{equation}
  k=4c, \, \Gamma^r{}_{\theta\theta}=0.  
\end{equation}
If $k=4c$, General B becomes
\begin{equation}
    \Gamma =\left(
\begin{array}{cccc}
 \{c,0,0,0\} & \{0,0,0,0\} & \left\{0,0,-\frac{1}{c},0\right\} & \left\{0,0,0,-\frac{\sin ^2\theta}{c}\right\} \\
 \left\{2 c^2 \Gamma^r{}_{\theta\theta},3 c,0,0\right\} & \left\{3 c,-\frac{(\Gamma^r{}_{\theta\theta})'-3}{\Gamma^r{}_{\theta\theta}},0,0\right\} & \{0,0,\Gamma^r{}_{\theta\theta},0\} & \left\{0,0,0,\sin ^2\theta \Gamma^r{}_{\theta\theta}\right\} \\
 \{0,0,c,0\} & \{0,0,0,0\} & \{c,0,0,0\} & \{0,0,0,-\sin \theta \cos \theta \} \\
 \{0,0,0,c\} & \{0,0,0,0\} & \{0,0,0,\cot \theta \} & \{c,0,\cot \theta ,0\} \\
\end{array}
\right).
\end{equation}
$\Gamma^r{}_{rr}=\infty$ when $\Gamma^r{}_{\theta\theta}=0$ so we should discard this solution.

As a result, the $\xi$ branch in $f(T)$ gravity has an EoMs correspondence in $f(Q)$ gravity, while 
the complex solution does not. This reveals that $f(T)$ solutions are not simply a subset of $f(Q)$ solutions with a complex solution beyond $f(Q)$ gravity in black hole background.

\begin{sidewaystable}
    \caption{EoMs-correspondence solutions, which are derived through solving Eq.~\eqref{eq:EoM_correspondence_eq}.}
    \vspace{10pt}
    \renewcommand{\arraystretch}{2}
    \begin{tabular}{ccccc}
    \hline
    \multirow{2}{*}{$\Gamma^r{}_{\theta\theta}$}& \multicolumn{2}{c}{\multirow{2}{*}{Branch}} & \multicolumn{2}{c}{$f(T)$}  \\
    \cmidrule(lr){4-5} 
    &\multicolumn{2}{c}{}&$\xi$&Complex\\
   \cmidrule(lr){1-5}
    \multirow{6}{*}{$f(Q)$}&\multirow{3}{*}{\small General A} & $E_{00}$ & $\xi \frac{r}{B}$ & $\pm i\frac{r}{B}$\\
    \cmidrule(lr){3-5}
    &&$E_{11}$&$\pm \frac{r}{B}$&$\pm \frac{r}{B}$ \\
    \cmidrule(lr){3-5}
    &&$E_{22}$&$\xi \frac{r}{B}$&$\pm \infty$ \\
    \cmidrule(lr){2-5}
    &\multirow{3}{*}{\small General B}
    &$E_{00}$&\small$\frac{\pm\sqrt{c r^2 A^2 (2 c-k) \left(c r^2 (2 c-k) (k-4 c)^2-2 k^2 A^2\right)}+4 c r A^2 (k-2 c)}{2 c \xi  B (2 c-k) \left(c r^2 (2 c-k)-2 A^2\right)}$ & \small $\pm\frac{i(4 c-k)}{\sqrt{4 c (2 c-k) \left(2 A^2-2 c^2 r^2+c k r^2\right)}}\frac{Ar}{B}$\\
    \cmidrule(lr){3-5}
    &&$E_{11}$&\small $\pm \frac{r A (4 c-k)}{2 \sqrt{c B^2 (2 c-k) \left(2 A^2+c r^2 (2 c-k)\right)}}$&\small $\pm \frac{(4 c-k)}{\sqrt{4 c (2 c-k) \left(2 A^2+2 c^2 r^2-c k r^2\right)}}
    \frac{Ar}{B}$\\
    \cmidrule(lr){3-5}
    &&$E_{22}$&\small$\frac{\pm\sqrt{-c^2 A^2 (k-2 c)^2 \left(r^2 (k-4 c)^2-4 A^2\right)}-2 c A^2  (k-2 c)}{2 c^2 \xi  r B(k-2 c)^2}$&\small $\pm \frac{i(4 c-k)}{2 c (2 c-k)}\frac{A}{B}$\\
    \hline
    \end{tabular}
    \label{tab:solutions_EoM_coorepondence}
\end{sidewaystable}

\subsection{Equations-of-motion  correspondence in cosmological  
spacetime}

Using Condition~\hyperref[cond1]{1}, we find only one branch that has the same 
non-metricity scalar value with the torsion scalar in $f(T)$ gravity:
\begin{align}
    {\Gamma^r}_{\theta\theta} &= -r, \quad &{\Gamma^r}_{\phi\phi} &= -r\sin^2\theta, \notag\\
    {\Gamma^\theta}_{r\theta} &= {\Gamma^\theta}_{\theta r} =\frac{1}{r}, \quad &{\Gamma^\theta}_{\phi\phi}&=-\cos\theta \sin\theta,\notag\\
    {\Gamma^\phi}_{r\phi} &= {\Gamma^\phi}_{\phi r}=\frac{1}{r}, \quad &{\Gamma^\phi}_{\theta\phi} &= {\Gamma^\phi}_{\phi \theta}=\cot\theta, \notag\\
    \Gamma^t{}_{tt} &=\gamma(t). & &
\end{align}
This branch  can be verified to yield the same equations of motion with $f(T)$ 
gravity, and thus we 
conclude that $f(T)$ solutions are a subset of $f(Q)$ solutions in 
cosmological  spacetimes. 

\section{Conclusions}
\label{sec:conclusion}

Metric-affine and tetrad-spin  formulations are generally considered to be 
equivalent descriptions of gravity. However, different constraints from the 
geometric background lead to distinct preferred formulations for various gravity 
theories. In this work, we have summarized the various theoretical branches that exist in torsional gravity and non-metric gravity. By comparing these branches, we have explored the correspondences between them. This analysis provides insight into how different branches of these gravitational theories can be related, paving the way for a deeper understanding of their mutual connections and potential unification. 

We have developed the tetrad-spin formulation of $f(Q)$ 
gravity to provide a novel perspective on STG. Based on the tetrad-spin 
formulation, we propose a Minkowski-equivalence correspondence between $f(Q)$ 
and $f(T)$ gravity. This correspondence is based on the equivalence between 
Lorentz vectors and St\"uckelberg fields, allowing us to establish an 
one-to-one mapping between certain solutions of $f(Q)$ and $f(T)$ gravity, which 
are obtained through switching off gravity.

The Minkowski-equivalence correspondence is derived from a vanishing curvature,  torsion and 
non-metricity tensor in Minkowski spacetime, which aligns naturally with 
physical intuition. However, symmetry analysis reveals additional solutions 
whose connections are not solely tied to coordinate transformations or Lorentz 
transformations. While these solutions are difficult to be interpreted, they 
cannot be dismissed from a mathematical perspective. In order to relate these 
general solutions, we propose another   correspondence, namely the 
equations-of-motion  correspondence, which is based on the equivalence of 
field equations in $f(Q)$ and $f(T)$ gravity. Despite the distinct geometrical 
perspectives of these two gravity theories, they can yield identical field 
equations under specific symmetry constraints. It is evident that $f(Q)$ gravity 
offers more flexibility in choosing the affine connection, resulting in a 
broader range of physical solutions compared to $f(T)$ gravity. Nevertheless, 
our analysis of EoMs correspondence reveals that the complex branch in $f(T)$ 
gravity lacks a corresponding solution in $f(Q)$ gravity in the black-hole background. In particular, the complex 
solution is derived from the tetrad-spin formulation, while in 
\cite{DAmbrosio:2021zpm}  the authors applied a metric-affine formulation to 
extract solutions of $f(T)$ gravity, resulting to   the real solutions only,
due to an inappropriate parameterization. This can be an example to 
explicitly show the importance of tetrad-spin formulation.

In summary, we observe numerous  similarities between $f(Q)$ and $f(T)$ 
gravity, which allow us to establish certain background-dependent 
correspondences between them. However, our current work focuses on the 
correspondences between the two theories at the background level. Moving 
forward, it will be crucial to investigate the correspondences and differences 
at the perturbative level, marking one direction for the extension of our 
current research. Additionally, to gain a deeper understanding of these 
correspondences, it is valuable to consider the entire framework from a more 
general perspective, namely, General Teleparallel Gravity (GTG), which is 
defined by the   absence of curvature only \cite{BeltranJimenez:2019odq, 
Heisenberg:2022mbo, Hohmann:2022mlc}. Developing the tetrad-spin formulation of 
GTG and clarifying the significance of these correspondences within this broader 
framework will be investigated in a future project.

\section*{Acknowledgments}
We are grateful to Yi-Fu Cai, Sebastian Bahamonde, Chunyu Li, Qingqing Wang for valuable discussions and insightful comments. 
This work is supported in part by the National Key R\&D Program of China 
(2021YFC2203100, 
2024YFC2207500), by NSFC (12433002, 12261131497, 92476203), by CAS young interdisciplinary innovation team (JCTD-2022-20), by 111 Project (B23042), by Anhui Postdoctoral Scientific Research Program
 Foundation (No. 2025C1184), by CSC Innovation Talent Funds, by USTC Fellowship for International Cooperation, 
and by USTC Research Funds of the Double First-Class Initiative. ENS 
acknowledges the contribution of the LISA 
CosWG, and of COST Actions  CA18108  ``Quantum Gravity Phenomenology in the multi-messenger approach''  and  CA21136 ``Addressing observational tensions in cosmology with systematics and fundamental physics (CosmoVerse)''.

\begin{appendices}
\section{Field equations of $f(Q)$ gravity in general tetrad-spin formulation
\label{appendix: EoMs}}

The action of $f(Q)$ gravity in general tetrad-spin formulation is: 
\begin{align}
        \mathcal{S}_{f(Q)}=-\frac{1}{2 \kappa}\int d^4x \, h f(Q)+\mathcal{S}_{matter}.
\end{align}
In order to perform the variation of this action, we utilize the following identities:
\begin{align}
    Q_{\lambda \mu \nu}&=-2A_{(\mu \nu )\lambda}= 
-2\eta_{ac}{h^c}_{(\mu}{h^b}_{\nu)}{A^a}_{b \lambda},\\
    \delta Q  &= \frac{1}{4}\delta(Q_{\alpha \mu  \nu}Q^{\alpha \mu 
\nu})-\frac{1}{2}\delta(Q_{\alpha \mu \nu}Q^{\mu \alpha 
\nu})-\frac{1}{4}\delta(Q_\alpha Q^\alpha)+\frac{1}{2}\delta(Q_\alpha 
\bar{Q}^\alpha).
\end{align}

\subsection{Field equations with respect to the tetrad \label{appendix: EoM with 
respect to the tetrad}}

Firstly, we calculate the variation of the non-metricity tensor: 
\begin{align}
    \delta_h Q_{\lambda \mu \nu}& = 2Q_{\lambda cb}{h^b}_{(\nu}\delta {h^c}_{\mu)},\\
    \delta_h Q_\alpha&=-4g^{\mu \nu}\eta_{a(c}{A^a}_{b)\alpha}{h^b}_{\nu} \delta{h^c}_{\mu}-2Q_{\alpha \mu \nu}g^{\mu \sigma}{h_a}^\nu \delta{h^a}_\sigma,\\
    \delta_h \bar{Q}_\alpha&=-2g^{\sigma(\mu}{h_a}^{\nu)}Q_{\mu \alpha \nu}\delta {h^a}_\sigma-4g^{\mu \nu}\eta_{a(c}{A^a}_{b)\mu}{h^b}_{(\nu}\delta{h^c}_{\alpha)}.
\end{align}
Next, we calculate the variation of the non-metricity scalar:
\begin{align}
    \delta_h(Q_{\alpha \mu \nu}Q^{\alpha \mu \nu})&=4Q^{\alpha \mu \nu}Q_{\alpha c \mu}\delta {h^c}_\nu-2(Q^{(\alpha}{}_{\mu \nu}Q^{\beta) \mu \nu}+2Q_{\rho}{}^{(\beta}{}_\nu Q^{|\rho|\alpha)\nu})\eta_{ab}{h^a}_\alpha \delta {h^b}_\beta,\\
    \delta_h (Q_{\alpha \mu \nu}Q^{\mu \alpha \nu})&= 4Q^{(\mu|\alpha|\nu)}Q_{\alpha c\mu}\delta{h^c}_\nu-2(2{Q^{(\alpha}}_{\mu \nu}Q^{|\mu |\beta) \nu}+{Q_{\rho \mu}}^{(\beta} Q^{|\mu \rho| \alpha)})\eta_{ab} {h^a}_\alpha\delta {h^b}_\beta,\\
    \delta_h(-\frac{1}{2}Q^\alpha Q_\alpha+Q_\alpha \bar{Q}^\alpha) 
    &=[(\bar{Q}^\alpha-Q^\alpha)g^{\mu \nu}+Q^{(\mu} g^{\nu)\alpha}](2Q_{\alpha c \nu }\delta{h^c}_\mu) +[(\bar{Q}^\alpha-Q^\alpha)Q_{\alpha \mu \nu}\notag\\
    &+Q^\alpha Q_{(\mu|\alpha|\nu) )}](-2g^{\mu \sigma}{h_a}^\nu \delta {h^a}_\sigma) +(Q^\alpha Q^\beta-2Q^{(\alpha} \bar{Q}^{\beta)})\eta_{ab}{h^b}_\beta \delta{h^a}_\alpha.
\end{align}
Combining these three terms  we obtain
\begin{align}
    \delta_h Q &= -2P^{\alpha \mu \nu}Q_{\alpha \rho \mu}h_c{}^\rho\delta{h^c}_\nu+(P_{(\nu|\alpha \rho|}{Q_{\mu)}}^{\alpha \rho}+2P^\alpha{}_{\rho(\mu}Q_{|\alpha|}{}^\rho{}_{\nu)}){h_a}^\mu g^{\nu \sigma}\delta {h^a}_\sigma.
\end{align}

\subsection{Field equations with respect to the spin connection}
Firstly, we calculate the variation of the spin connection:
\begin{align}
    \delta_A Q_{\lambda \mu \nu}&=-2\eta_{ac}{h^c}_{(\mu}{h^b}_{\nu)}\delta{A^a}_{b \lambda},\\
    \delta_A Q_\alpha&=-2 \delta{{A}^a}_{a \alpha},\\
    \delta_A \bar{Q}_\alpha &=-2\eta_{ac}g^{\mu \nu}{h^c}_{(\alpha}{h^b}_{\nu)}\delta {A^a}_{b \mu}.
\end{align}
The variation of the non-metricity scalar is given by:
\begin{align}
    \delta_A(Q_{\alpha \mu \nu}Q^{\alpha \mu \nu})&= -4{{Q^\alpha}_a}^b \delta{A^a}_{b \alpha},\\
    \delta_A(Q_{\alpha \mu \nu}Q^{\mu \alpha \nu})&= -4\eta_{ac}Q^{(c|\alpha|b)}\delta{A^a}_{b \alpha},\\
    \delta_A(-\frac{1}{2}Q^\alpha Q_\alpha+Q_\alpha \bar{Q}^\alpha) &=-2[{\delta^a}_b(\bar{Q}^\mu-Q^\mu)+\eta_{ac}g^{\mu (\nu}Q^{\alpha)}{h^c}_{\alpha}{h^b}_{\nu}]\delta {A^a}_{b \mu}.
\end{align}
Adding these three terms   yields:
\begin{align}
    \delta_A Q=2{{P^\mu}_a}^b \delta{A^a}_{b \mu}.
\end{align}

\subsection{Decomposing the variation of the spin connection}

 Recalling Eq.~\eqref{eq:spin_general_expression}, we decompose the variation of the spin 
connection as:
\begin{align}
    \delta_A A^a{}_{b \mu} &= \delta_h A^a{}_{b \mu} + \delta_\xi A^a{}_{b \mu}.
    \label{spinconsplit}
\end{align}
For the first term, through direct calculation  we obtain 
\begin{equation}
    \delta_h A^a{}_{b \mu}=A^c{}_{b \mu}h_c{}^\rho \delta h^a{}_\rho-A^a{}_{c \mu}h_b{}^\rho\delta h^c{}_\rho-\partial_\mu (h_b{}^\sigma \delta h^a{}_\sigma).
\end{equation}
Inserting the above expression into the Lagrangian, we acquire
\begin{align}
    hf_QP^\mu{}_a{}^b \delta_h A^a{}_{b 
    \mu}=&hf_QP^\mu{}_a{}^b(A^c{}_{b \mu}h_c{}^\rho \delta h^a{}_\rho-A^a{}_{c 
    \mu}h_b{}^\rho\delta h^c{}_\rho-\partial_\mu (h_b{}^\sigma \delta h^a{}_\sigma)) 
    \notag\\
    =&\partial_\mu(hf_QP^\mu{}_a{}^b)h_b{}^\sigma\delta h^a{}_\sigma+hf_Q(P^\mu{}_a{}^bA^c{}_{b \mu}h_c{}^\rho-P^\mu{}_c{}^bA^c{}_{a \mu}h_b{}^\rho)\delta h^a{}_\rho \notag\\
    =&\mathring{\nabla}_\mu(hf_QP^\mu{}_a{}^b)h_b{}^\sigma\delta h^a{}_\sigma+\Gamma^\nu{}_{\nu \mu}hf_QP^\mu{}_a{}^bh_b{}^\sigma\delta h^a{}_\sigma-\Gamma^\mu{}_{\nu \mu}hf_QP^\nu{}_a{}^bh_b{}^\sigma\delta h^a{}_\sigma,
\end{align}
where     $\mathring{\nabla}_\mu$  denotes the covariant derivative with 
respect to both the coordinate and tangent indices: $\mathring{\nabla}_\mu 
V^a{}_\nu \equiv \partial_\mu V^a{}_\nu + A^a{}_{b 
\mu}V^b{}_\nu-\Gamma^\rho{}_{\nu \mu}V^a{}_\rho$.
In STG, since the torsion tensor is zero, we obtain
\begin{align}
    hf_QP^\mu{}_a{}^b \delta_h A^a{}_{b \mu}=\mathring{\nabla}_\mu(hf_QP^\mu{}_a{}^b)h_b{}^\sigma\delta h^a{}_\sigma.
\end{align}
Having in mind  Appendix~\ref{appendix: EoM with respect to the tetrad}, 
the field equations in terms of the tetrad are expressed as
\begin{align}   
&\frac{h_a{}^\rho}{h}\nabla_\nu(hf_QP^\nu{}_\rho{}^\mu)+\frac{1}{2}f_Q[-2{h_a}
^\rho Q_{\alpha \rho \nu}P^{\alpha \nu \mu}+{h_a}^\alpha g^{\beta 
\mu}(P_{(\beta|\nu \rho}{Q_{\alpha)}}^{\nu 
\rho}+2P^\nu{}_{\rho(\alpha}Q_{|\nu|}{}^{\rho}{}_{\beta)})] \nonumber \\
 &+\frac{1}{2}f\,{h_a}^\mu
    =\kappa {\mathcal{T}_a}^\mu,
\end{align}
where we have used the tetrad postulate Eq.~\eqref{eq: tetrad postulate} in 
order to simplify their form. 
By further simplification, we can arrive at 
\begin{align}
    &\frac{g_{\sigma \rho}h_a{}^\rho}{h}\nabla_\nu(hf_QP^{\nu\sigma\mu})+\frac{1}{2}f_Q{h_a}^\alpha g^{\beta \mu}(P_{(\beta|\nu \rho}{Q_{\alpha)}}^{\nu \rho}+2P^\nu{}_{\rho(\alpha}Q_{|\nu|}{}^{\rho}{}_{\beta)})+\frac{1}{2}f\,{h_a}^\mu=\kappa {\mathcal{T}_a}^\mu.
\end{align}
We mention that this is the same as Eq.~\eqref{eq: f(Q) metric EoM} in the 
metric-affine formulation.

For the second term in Eq.~\eqref{spinconsplit} we utilize the following equations
\begin{align}
    \delta \frac{\partial \xi^\alpha}{\partial x^\mu} &=  
\frac{\partial}{\partial x^\mu} \delta \xi^\alpha,\\
\delta \frac{\partial x^\mu}{\partial \xi^\alpha} &=  -\frac{\partial 
x^\mu}{\partial \xi^\beta}\frac{\partial x^\nu}{\partial 
\xi^\alpha}\frac{\partial}{\partial x^\nu}\delta \xi^\beta,
\end{align}
to calculate the variation of the spin connection with respect to the St\"uckelberg fields as
\begin{align}
    \delta_\xi A^a{}_{b \mu} =& h^a{}_\rho\frac{\partial x^\rho}{\partial 
\xi^\alpha}  \frac{\partial}{\partial x^\mu}[h_b{}^\nu(\frac{\partial \delta 
\xi^\alpha}{\partial x^\nu})]-h^a{}_\rho \partial_\mu(\frac{\partial 
\xi^\alpha}{\partial x^\nu}h_b{}^\nu)\frac{\partial x^\rho}{\partial 
\xi^\beta}\frac{\partial x^\sigma
    }{\partial \xi^\alpha}\frac{\partial}{\partial x^\sigma}\delta \xi^\beta\\ 
    =&h^a{}_\rho\frac{\partial x^\rho}{\partial \xi^\alpha}h_b{}^\nu\partial_\mu 
 \partial_\nu\delta \xi^\alpha+(h^a{}_\rho \partial_\mu h_b{}^\nu-h^a{}_\rho 
h_c{}^{\nu} A^c{}_{b \mu})\frac{\partial x^\rho}{\partial 
\xi^\alpha}\partial_\nu \delta \xi^\alpha \label{A.30}\\
    =&h^a{}_\rho\frac{\partial x^\rho}{\partial \xi^\alpha}h_b{}^\nu\partial_\mu 
 \partial_\nu\delta \xi^\alpha+h^a{}_\rho \mathcal{D}_\mu 
h_b{}^\nu\frac{\partial x^\rho}{\partial \xi^\alpha}\partial_\nu \delta 
\xi^\alpha \label{A.31}\\
    =&h^a{}_\rho\frac{\partial x^\rho}{\partial \xi^\alpha}h_b{}^\nu\partial_\mu 
 \partial_\nu\delta \xi^\alpha-h^a{}_\rho h_b{}^\sigma \Gamma^\nu{}_{\sigma 
\mu}\frac{\partial x^\rho}{\partial \xi^\alpha}\partial_\nu \delta \xi^\alpha 
\label{A.32}\\
    =&h^a{}_\rho h_b{}^\nu \frac{\partial x^\rho}{\partial \xi^\alpha} 
\nabla_\mu(\partial_\nu \delta \xi^\alpha) \label{A.33}\\
    =&h^a{}_\rho h_b{}^\nu \frac{\partial x^\rho}{\partial \xi^\alpha} 
\nabla_\mu\nabla_\nu \delta \xi^\alpha \label{A.34}.
\end{align}
For clarity we mention that from  Eq.~\eqref{A.30} to Eq.~\eqref{A.31}  we 
applied the definition Eq.~\eqref{2.5}, from  Eq.~\eqref{A.31} to Eq.~\eqref{A.32}  we utilized the definition  Eq.~\eqref{eq: affine connection 
definition}, and from Eq.~\eqref{A.32} to Eq.~\eqref{A.34}  we employed the 
property of St\"uckelberg fields that they are invariant under the coordinate 
transformation.

\section{The general affine connection in flat, torsion-free,  static and 
spherically symmetric spacetime through metric-affine approach \label{appendix:metric_afffine_to_general_A}}

In \cite{Hohmann:2019fvf}, the authors derive the general form of the  affine 
connection in flat and spherically symmetric metric-affine geometry. In the 
static case, the form can be expressed as
\begin{align}
    \Gamma^t{}_{tr}&=\frac{F_1'}{F_1}+F_3' \tanh (F_3-F_4), &\Gamma^t{}_{rr}&=\frac{F_2 F_4' \text{sech}(F_3-F_4)}{F_1},\notag\\
    \Gamma^t{}_{\theta \theta}&= \frac{F_5 \sinh (F_4) \cos (F_6) \text{sech}(F_3-F_4)}{F_1}, &\Gamma^t{}_{\theta \phi}&=\Gamma^t{}_{\theta\theta}\tan(F_4)\sin(\theta), \notag\\
    \Gamma^t{}_{\phi \theta}&=-\Gamma^t{}_{\theta \phi}, &\Gamma^t{}_{\phi \phi}&=\Gamma^t{}_{\theta\theta}\sin^2(\theta);\notag\\
    \Gamma^r{}_{t r}&= \frac{F_1 F_3' \text{sech}(F_3-F_4)}{F_2} , &\Gamma^r{}_{r r}&=\frac{F_2'}{F_2}-F_4' \tanh (F_3-F_4),\notag\\
    \Gamma^r{}_{\theta \theta}&=  -\frac{F_5 \cosh (F_3) \cos (F_6) 
\text{sech}(F_3-F_4)}{F_2}, &\Gamma^r{}_{\theta 
\phi}&=\Gamma^r{}_{\theta\theta}\tan(F_6)\sin(\theta), \notag\\
    \Gamma^r{}_{\phi \theta}&=-\Gamma^r{}_{\theta \phi}, &\Gamma^r{}_{\phi 
\phi}&=\Gamma^r{}_{\theta\theta}\sin^2(\theta);\notag\\ 
    \Gamma^\theta{}_{t \theta}&=  \frac{F_1 \sinh (F_3) \cos (F_6)}{F_5}, 
&\Gamma^\theta{}_{t 
\phi}&=\Gamma^\theta{}_{t\theta}\tan(F_6)\sin(\theta),\notag\\ 
    \Gamma^\theta{}_{r \theta}&= \frac{F_2 \cosh (F_4) \cos (F_6)}{F_5}, 
&\Gamma^\theta{}_{r 
\phi}&=\Gamma^\theta{}_{r\theta}\tan(F_6)\sin(\theta),\notag\\ 
    \Gamma^\theta{}_{\theta r}&= \frac{F_5'}{F_5}, &\Gamma^\theta{}_{\phi r}&= 
-\sin (\theta ) F_6',\notag
\\
    \Gamma^\theta{}_{\phi \phi}&= -\sin (\theta )\cos (\theta );&&\notag\\
    \Gamma^\phi{}_{t\theta}&= -\frac{F_1 \csc (\theta ) \sinh (F_3) \sin (F_6)}{F_5}, &\Gamma^\phi{}_{t \phi}&=-\Gamma^\phi{}_{t\theta}\cot(F_6)\sin(\theta),\notag\\
    \Gamma^\phi{}_{r \theta}&= -\frac{F_2 \csc (\theta ) \cosh (F_4) \sin (F_6)}{F_5}, &\Gamma^\phi{}_{r \phi}&=-\Gamma^\phi{}_{r\theta}\cot(F_6)\sin(\theta),\notag\\
    \Gamma^\phi{}_{\theta r}&=  \csc (\theta ) F_6', &\Gamma^\phi{}_{\theta \phi}&= \cot (\theta ),\notag\\
    \Gamma^\phi{}_{\phi r}&=  \frac{\text{F}_5'}{F_5}, &\Gamma^\phi{}_{\phi \theta}&= \cot (\theta ),
    \label{eq:general_affine_con_f(Q)_ss}
\end{align}
where $\{F_i(r)\}$ ($i=1,2,3,4,5,6$) are functions of $r$.

We can also choose an alternative set of parameters, defined as
\begin{align}
    C_1&=F_1 \cosh(F_3), &C_3&=F_1\sinh(F_3), \notag\\
    C_2&=F_2\sinh(F_4), &C_4&=F_2\cosh(F_4),\notag\\
    C_5&=F_5\cos(F_6), &C_6&=F_5\sin(F_6).
\end{align}
The connection can be rewritten as
\begin{align}
    \Gamma^t{}_{tr}&= \frac{C_4 C_1'-C_2 C_3'}{C_1 C_4-C_2 C_3}, &\Gamma^t{}_{rr}&=\frac{C_4 C_2'-C_2 C_4'}{C_1 C_4-C_2 C_3},\notag\\
    \Gamma^t{}_{\theta \theta}&= \frac{C_2 C_5}{C_1 C_4-C_2 C_3} , &\Gamma^t{}_{\theta \phi}&=\frac{C_2 C_6 \sin (\theta )}{C_1 C_4-C_2 C_3}, \notag\\
    \Gamma^t{}_{\phi \theta}&=-\Gamma^t{}_{\theta \phi}, &\Gamma^t{}_{\phi \phi}&=\Gamma^t{}_{\theta\theta}\sin^2(\theta);\notag\\
    \Gamma^r{}_{t r}&= \frac{C_3 C_1'-C_1 C_3'}{C_2 C_3-C_1 C_4} , &\Gamma^r{}_{r r}&=\frac{C_3 C_2'-C_1 C_4'}{C_2 C_3-C_1 C_4},\notag\\
    \Gamma^r{}_{\theta \theta}&=  \frac{C_1 C_5}{C_2 C_3-C_1 C_4} , &\Gamma^r{}_{\theta \phi}&=\frac{C_1 C_6 \sin (\theta )}{C_2 C_3-C_1 C_4} , \notag\\
    \Gamma^r{}_{\phi \theta}&=-\Gamma^r{}_{\theta \phi}, &\Gamma^r{}_{\phi \phi}&=\Gamma^r{}_{\theta\theta}\sin^2(\theta);\notag\\
    \Gamma^\theta{}_{t \theta}&= \frac{C_3 C_5}{C_5^2+C_6^2}, &\Gamma^\theta{}_{t \phi}&=\frac{C_3 C_6 \sin (\theta )}{C_5^2+C_6^2},\notag\\
    \Gamma^\theta{}_{r \theta}&= \frac{C_4 C_5}{C_5^2+C_6^2}, &\Gamma^\theta{}_{r \phi}&= \frac{C_4 C_6 \sin (\theta )}{C_5^2+C_6^2},\notag\\
    \Gamma^\theta{}_{\theta r}&= \frac{C_5 C_5'+C_6 C_6'}{C_5^2+C_6^2}, &\Gamma^\theta{}_{\phi r}&= \frac{\sin (\theta ) \left(C_6 C_5'-C_5 C_6'\right)}{C_5^2+C_6^2},\notag\\
    \Gamma^\theta{}_{\phi \phi}&= -\sin (\theta )\cos (\theta );&&\notag\\
    \Gamma^\phi{}_{t\theta}&=  -\frac{C_3 C_6 \csc (\theta )}{C_5^2+C_6^2}, &\Gamma^\phi{}_{t \phi}&=\frac{C_3 C_5}{C_5^2+C_6^2},\notag\\
    \Gamma^\phi{}_{r \theta}&= -\frac{C_4 C_6 \csc (\theta )}{C_5^2+C_6^2}, &\Gamma^\phi{}_{r \phi}&= \frac{C_4 C_5}{C_5^2+C_6^2},\notag\\
    \Gamma^\phi{}_{\theta r}&= \frac{\csc (\theta ) \left(C_5 C_6'-C_6 C_5'\right)}{C_5^2+C_6^2}, &\Gamma^\phi{}_{\theta \phi}&= \cot (\theta ),\notag\\
    \Gamma^\phi{}_{\phi r}&= \frac{C_5 C_5'+C_6 C_6'}{C_5^2+C_6^2}, &\Gamma^\phi{}_{\phi \theta}&= \cot (\theta ).
\end{align}
The torsion tensor is calculated as
\begin{align}
    T^t{}_{tr}&=-T^t{}_{rt}=\frac{C_4 C_1'-C_2 C_3'}{C_2 C_3-C_1 C_4}, &T^t{}_{\theta \phi}&=-T^t{}_{\theta \phi}=\frac{2 C_2 C_6 \sin (\theta )}{C_2 C_3-C_1 C_4},\notag\\
    T^r{}_{tr}&=-T^r{}_{rt}=\frac{C_3 C_1'-C_1 C_3'}{C_1 C_4-C_2 C_3}, &T^r{}_{\theta \phi}&=-T^r{}_{\theta \phi}=\frac{2 C_1 C_6 \sin (\theta )}{C_1 C_4-C_2 C_3}, \notag\\
    T^\theta{}_{t\theta}&=-T^\theta{}_{\theta t}=-\frac{C_3 C_5}{C_5^2+C_6^2}, &T^\theta{}_{t\phi}&=-T^\theta{}_{\phi t}=-\frac{C_3 C_6 \sin (\theta )}{C_5^2+C_6^2}, \notag\\
    T^\theta{}_{r\theta}&=-T^\theta{}_{\theta r}=\frac{-C_4 C_5+C_5 C_5'+C_6 C_6'}{C_5^2+C_6^2}, &T^\theta{}_{r\phi}&=T^\theta{}_{\phi r}=-\frac{\sin (\theta ) \left(C_4 C_6-C_6 C_5'+C_5 C_6'\right)}{C_5^2+C_6^2}, \notag\\
    T^\phi{}_{t\theta}&=-T^\phi{}_{\theta t}=\frac{C_3 C_6 \csc (\theta )}{C_5^2+C_6^2}, &T^\phi{}_{t\phi}&=-T^\phi{}_{\phi t}=-\frac{C_3 C_5}{C_5^2+C_6^2},\notag\\
    T^\phi{}_{r\theta}&=-T^\phi{}_{\theta r}=\frac{\csc (\theta ) \left(C_4 C_6-C_6 C_5'+C_5 C_6'\right)}{C_5^2+C_6^2},& T^\phi{}_{r\phi}&=T^\phi{}_{\phi r}=\frac{-C_4 C_5+C_5 C_5'+C_6 C_6'}{C_5^2+C_6^2}.
\end{align}
Note that   solutions satisfying the  torsion-free condition are
\begin{align}
    C_1(r)&=k_1 \ne 0,\\
    C_3(r)&=C_6(r)=0,\\
    C_4(r)&=C_5'(r),
\end{align}
where $k_1$ is a constant. 
By substituting those solutions into the connection, we ultimately obtain Eq.~\eqref{eq: affine connection in SS}. 

According to the metric field equations  Eq.~\eqref{eq: f(Q) metric EoM}, the 
off-diagonal components of the field equations vanish, while the diagonal 
components are \label{appendix:EoMs_metric_affine_to_general_A}
\begin{align}
    E_{00}&=-\frac{1}{2} A^2 f \notag\\
    &\quad + \frac{A}{2 r^2 B^3 C_5^2 C_5'^2} f_Q \Big
    \{2 B C_5 A' C_5' \left[B^2 
C_5^2+r C_5' \left(r C_5'-2 C_5\right)\right]-2 r A C_5 B' \left(r C_5'-2 
C_5\right) C_5'^2 \notag\\
    & \quad \quad+2 A B^2 C_5^3 B' C_5' -2 A B^3 C_5^3C_5''-2 r^2 A B C_5'^4+4 r 
A B C_5 C_5'^3+2 A B^3 C_5^2 C_5'^2
\notag\\
    & \quad \quad
    -4 A B C_5^2 C_5'^2+2 r^2 A B C_5 C_5'^2 C_5''\Big\}\notag \\
    & \quad +\frac{A^2 f_{QQ}Q' \left(2 r^2 C_5'^2 +2  B^2 C_5^2  -4 r C_5 C_5' 
\right)}{2 r^2 B^2 C_5 C_5'},\\
    E_{11} &=\frac{1}{2} B^2 f \notag\\
    & \quad + \frac{f_Q}{2 r^2 A B C_5^2 C_5'^2}\Big\{-2 B C_5 A' C_5' 
\left[B^2 C_5^2+r C_5' \left(r C_5'-4 C_5\right)\right]+2 r^2 A C_5 B' C_5'^3-2 
A B^2 C_5^3 B' C_5'\notag\\
    &\quad \quad+2 A B^3 C_5^3C_5''+2 r^2 A B C_5'^4-4 r A B C_5 C_5'^3-2 A B^3 
C_5^2 C_5'^2+4 A B C_5^2 C_5'^2-2 r^2 A B C_5 C_5'^2C_5''\Big\} \notag\\
    & \quad +\frac{f_{QQ}Q' \left(2 r^2 C_5'^2 -2  B^2 C_5^2\right)}{2 r^2   C_5 C_5'},\\
     E_{22} &=\frac{E_{33}}{\sin^2(\theta)} =\frac{1}{2} r^2 f \notag\\
    & \quad +\frac{f_Q}{2 A B^3 C_5^2 C_5'^2} \Big[2 r^2 B C_5^2 A''(r) 
C_5'^2-2 r^2 C_5^2 A' B' C_5'^2-2 r^2 B C_5 A' C_5'^3+6 r B C_5^2 A' 
C_5'^2\notag\\
    & \quad \quad-2 B^3 C_5^3 A' C_5'+2 r A C_5 B' \left(r C_5'-C_5\right) 
C_5'^2-2 A B^2 C_5^3 B' C_5' +2 A B^3 C_5^3C_5''+2 r^2 A B C_5'^4\notag\\
    & \quad \quad -4 r A B C_5 C_5'^3+2 A B C_5^2 C_5'^2-2 r^2 A B C_5 
C_5'^2C_5''\Big]\notag\\
    & \quad +\frac{f_{QQ}Q' \left(2 r^2 C_5 A' -2 r^2 A C_5' +2 r A C_5\right)}{2 A B^2 C_5 }.
\end{align}

\section{Equations of motions for the general affine connections of $f(Q)$ gravity in static and spherically symmetric spacetime \label{appendix:EoMs_general_affine_connection}}

For the general affine connection A~\eqref{eq:f(Q)_general_connection_A}, EoMs are
\begin{align}
    E_{00}=
    &-\frac{A \left(2 r^2 A B \Gamma^r{}_{\theta\theta}+2 A B^3 (\Gamma^r{}_{\theta\theta})^3+4 r A B (\Gamma^r{}_{\theta\theta})^2\right)}{2 r^2 B^3 (\Gamma^r{}_{\theta\theta})^2}Q'f_{QQ}\notag\\
    &-\frac{A}{2 r^2 B^3 (\Gamma^r{}_{\theta\theta})^2} [2 B \Gamma^r{}_{\theta\theta} A' \left(B^2 (\Gamma^r{}_{\theta\theta})^2+r^2+2 r \Gamma^r{}_{\theta\theta}\right)+2 A B^2 (\Gamma^r{}_{\theta\theta})^3 B' \notag\\
    &-2 r A \Gamma^r{}_{\theta\theta} (2 \Gamma^r{}_{\theta\theta}+r) B'-2 r^2 A B (\Gamma^r{}_{\theta\theta})'+2 A B^3 (\Gamma^r{}_{\theta\theta})^2 (\Gamma^r{}_{\theta\theta})'+4 A B (\Gamma^r{}_{\theta\theta})^2+4 r A B \Gamma^r{}_{\theta\theta}]f_Q\notag\\
    &-\frac{1}{2} A^2 f,\\
    E_{11}=
    &\frac{\left(2 A B^3 (\Gamma^r{}_{\theta\theta})^3-2 r^2 A B \Gamma^r{}_{\theta\theta}\right)}{2 r^2 A B (\Gamma^r{}_{\theta\theta})^2}Q'f_{QQ}\notag\\
    &+\frac{1}{2 r^2 A B (\Gamma^r{}_{\theta\theta})^2} [2 B \Gamma^r{}_{\theta\theta} A' \left(B^2 (\Gamma^r{}_{\theta\theta})^2+r^2+4 r \Gamma^r{}_{\theta\theta}\right)-2 r^2 A \Gamma^r{}_{\theta\theta} B'+2 A B^2 (\Gamma^r{}_{\theta\theta})^3 B'\notag\\
    &-2 r^2 A B (\Gamma^r{}_{\theta\theta})'+2 A B^3 (\Gamma^r{}_{\theta\theta})^2 (\Gamma^r{}_{\theta\theta})'+4 A B (\Gamma^r{}_{\theta\theta})^2+4 r A B \Gamma^r{}_{\theta\theta}]f_Q\notag\\
    &+\frac{1}{2} B^2 f,\\
    E_{22}=&\frac{E_{33}}{\sin^2\theta}=\frac{\left(2 r^2 B (\Gamma^r{}_{\theta\theta})^2 A'+2 r^2 A B \Gamma^r{}_{\theta\theta}+2 r A B (\Gamma^r{}_{\theta\theta})^2\right)}{2 A B^3 (\Gamma^r{}_{\theta\theta})^2} Q'f_{QQ} \notag\\
    &+\frac{1}{2 A B^3 (\Gamma^r{}_{\theta\theta})^2} [2 r^2 B (\Gamma^r{}_{\theta\theta})^2 A''(r)-2 r^2 (\Gamma^r{}_{\theta\theta})^2 A' B'+2 B^3 (\Gamma^r{}_{\theta\theta})^3 A'+2 r B \Gamma^r{}_{\theta\theta} (3 \Gamma^r{}_{\theta\theta}+r) A'\notag\\
    &+2 A B^2 (\Gamma^r{}_{\theta\theta})^3 B'-2 r A \Gamma^r{}_{\theta\theta} (\Gamma^r{}_{\theta\theta}+r) B'-2 r^2 A B (\Gamma^r{}_{\theta\theta})'+2 A B^3 (\Gamma^r{}_{\theta\theta})^2 \left((\Gamma^r{}_{\theta\theta})'+1\right)\notag\\
    &+2 A B (\Gamma^r{}_{\theta\theta})^2+4 r A B \Gamma^r{}_{\theta\theta}]f_Q\notag\\
    &+\frac{1}{2} r^2 f.
\end{align}

For the general affine connection B~\eqref{eq:f(Q)_general_connection_B}, EoMs are
\begin{align}
    E_{00}=
    &\frac{1}{8 c r^2 A B^3 (2 c-k) (\Gamma^r{}_{\theta\theta})^2}[4 c^2 r^2 A B^3 (k-2 c)^2 (\Gamma^r{}_{\theta\theta})^3-r^2 A^3 B (k-4 c)^2 \Gamma^r{}_{\theta\theta}\notag\\
    &-8 c A^3 B^3 (2 c-k) (\Gamma^r{}_{\theta\theta})^3-16 c r A^3 B (2 c-k) (\Gamma^r{}_{\theta\theta})^2] Q' f_{QQ}\notag\\
    &+\frac{1}{8 c r^2 A B^3 (2 c-k) (\Gamma^r{}_{\theta\theta})^2} [4 c^2 r^2 B^3 (k-2 c)^2 (\Gamma^r{}_{\theta\theta})^3 A'\notag\\
    &-A^2 B \Gamma^r{}_{\theta\theta} A' \left(8 c B^2 (2 c-k) (\Gamma^r{}_{\theta\theta})^2+r^2 (k-4 c)^2+16 c r (2 c-k) \Gamma^r{}_{\theta\theta}\right)\notag\\
    &-4 c^2 r^2 A B^2 (k-2 c)^2 (\Gamma^r{}_{\theta\theta})^3 B'+8 c A^3 B^2 (k-2 c) (\Gamma^r{}_{\theta\theta})^3 B'\notag\\
    &+r A^3 \Gamma^r{}_{\theta\theta} B' \left(16 c (2 c-k) \Gamma^r{}_{\theta\theta}+r (k-4 c)^2\right)-4 c^2 r^2 A B^3 (k-2 c)^2 (\Gamma^r{}_{\theta\theta})^2 (\Gamma^r{}_{\theta\theta})'\notag\\
    &-8 c^2 r A B^3 (k-2 c)^2 (\Gamma^r{}_{\theta\theta})^3+r^2 A^3 B (k-4 c)^2 (\Gamma^r{}_{\theta\theta})'-8 c A^3 B^3 (2 c-k) (\Gamma^r{}_{\theta\theta})^2 (\Gamma^r{}_{\theta\theta})'\notag\\
    &-16 c A^3 B (2 c-k) (\Gamma^r{}_{\theta\theta})^2-2 r A^3 B (k-4 c)^2 \Gamma^r{}_{\theta\theta}]f_Q 
    \notag\\
    &-\frac{1}{2} A^2 f,\\
    E_{11}=
    &\frac{1} {8 c r^2 A^3 B (2 c-k) (\Gamma^r{}_{\theta\theta})^2} [4 c^2 r^2 A B^3 (k-2 c)^2 (\Gamma^r{}_{\theta\theta})^3-r^2 A^3 B (k-4 c)^2 \Gamma^r{}_{\theta\theta}\notag\\
    &+8 c A^3 B^3 (2 c-k) (\Gamma^r{}_{\theta\theta})^3]Q' f_{QQ} \notag\\
    &+\frac{1} {8 c r^2 A^3 B (2 c-k) (\Gamma^r{}_{\theta\theta})^2}[-4 c^2 r^2 B^3 (k-2 c)^2 (\Gamma^r{}_{\theta\theta})^3 A'\notag\\
    &+A^2 B \Gamma^r{}_{\theta\theta} A' \left(8 c B^2 (2 c-k) (\Gamma^r{}_{\theta\theta})^2+r^2 (k-4 c)^2+32 c r (2 c-k) \Gamma^r{}_{\theta\theta}\right)\notag\\
    &+4 c^2 r^2 A B^2 (k-2 c)^2 (\Gamma^r{}_{\theta\theta})^3 B'-r^2 A^3 (k-4 c)^2 \Gamma^r{}_{\theta\theta} B'\notag\\
    &+8 c A^3 B^2 (2 c-k) (\Gamma^r{}_{\theta\theta})^3 B'+4 c^2 r^2 A B^3 (k-2 c)^2 (\Gamma^r{}_{\theta\theta})^2 (\Gamma^r{}_{\theta\theta})'\notag\\
    &+8 c^2 r A B^3 (k-2 c)^2 (\Gamma^r{}_{\theta\theta})^3-r^2 A^3 B (k-4 c)^2 (\Gamma^r{}_{\theta\theta})'+8 c A^3 B^3 (2 c-k) (\Gamma^r{}_{\theta\theta})^2 (\Gamma^r{}_{\theta\theta})'\notag\\
    &-16 c A^3 B (k-2 c) (\Gamma^r{}_{\theta\theta})^2+2 r A^3 B (k-4 c)^2 \Gamma^r{}_{\theta\theta}]f_Q 
    \notag\\
    &+\frac{1}{2} B^2 f,\\
    E_{22}=&
    \frac{E_{33}}{\sin^2\theta}=\frac{1}{8 c A^3 B^3 (2 c-k) (\Gamma^r{}_{\theta\theta})^2} [8 c r^2 A^2 B (2 c-k) (\Gamma^r{}_{\theta\theta})^2 A'+4 c^2 r^2 A B^3 (k-2 c)^2 (\Gamma^r{}_{\theta\theta})^3
    \notag\\
    &+r^2 A^3 B (k-4 c)^2 \Gamma^r{}_{\theta\theta}+8 c r A^3 B (2 c-k) (\Gamma^r{}_{\theta\theta})^2]Q' f_{QQ} 
    \notag\\
    &+\frac{1}{8 c A^3 B^3 (2 c-k) (\Gamma^r{}_{\theta\theta})^2} [8 c r^2 A^2 B (2 c-k) (\Gamma^r{}_{\theta\theta})^2 A''(r)+8 c r^2 A^2 (k-2 c) (\Gamma^r{}_{\theta\theta})^2 A' B'
    \notag\\
    &-4 c^2 r^2 B^3 (k-2 c)^2 (\Gamma^r{}_{\theta\theta})^3 A'+8 c A^2 B^3 (2 c-k) (\Gamma^r{}_{\theta\theta})^3 A'
    \notag\\
    &+r A^2 B \Gamma^r{}_{\theta\theta} A' \left(24 c (2 c-k) \Gamma^r{}_{\theta\theta}+r (k-4 c)^2\right)+4 c^2 r^2 A B^2 (k-2 c)^2 (\Gamma^r{}_{\theta\theta})^3 B'
    \notag\\
    &+8 c A^3 B^2 (2 c-k) (\Gamma^r{}_{\theta\theta})^3 B'-r A^3 \Gamma^r{}_{\theta\theta} B' \left(8 c (2 c-k) \Gamma^r{}_{\theta\theta}+r (k-4 c)^2\right)
    \notag\\
    &+4 c^2 r^2 A B^3 (k-2 c)^2 (\Gamma^r{}_{\theta\theta})^2 (\Gamma^r{}_{\theta\theta})'+8 c^2 r A B^3 (k-2 c)^2 (\Gamma^r{}_{\theta\theta})^3-r^2 A^3 B (k-4 c)^2 (\Gamma^r{}_{\theta\theta})'
    \notag\\
    &+8 c A^3 B^3 (2 c-k) (\Gamma^r{}_{\theta\theta})^2 \left((\Gamma^r{}_{\theta\theta})'+1\right)+8 c A^3 B (2 c-k) (\Gamma^r{}_{\theta\theta})^2+2 r A^3 B (k-4 c)^2 \Gamma^r{}_{\theta\theta}]f_Q
    \notag\\
    &+\frac{1}{2} r^2 f.
\end{align}
\end{appendices}


\bibliography{sn-bibliography}

\end{document}